\documentclass[12pt]{article}

\usepackage[utf8]{inputenc}
\usepackage[english]{babel}

\usepackage{xcolor}
\usepackage{amsmath}
\usepackage{amsthm}
\usepackage{amssymb}
\usepackage[left=3cm,right=3cm,top=3.5cm,bottom=3.5cm,bindingoffset=0cm]{geometry}
\usepackage{dsfont}
\usepackage{bm}
\usepackage{leftidx}
\usepackage[makeroom]{cancel}
\usepackage{indentfirst}
\usepackage{cite}
\usepackage{caption}
\usepackage{hyperref}
\usepackage{tikz}
\usepackage{subfig}
\usepackage{graphicx}
\usepackage{tikz-cd}
\usepackage[titletoc,title]{appendix}

\usepackage{adjustbox}
\usepackage{multirow}
\usetikzlibrary{matrix,arrows,decorations.pathmorphing}
\newcommand {\unit} {\mathds{1}}

\DeclareMathOperator{\sn}{sn\s{}  }
\DeclareMathOperator{\snt}{sn^2\s{}  }

\DeclareMathOperator{\cnt}{cn^2\s{}  }

\DeclareMathOperator{\diag}{diag\s{}  }
\DeclareMathOperator{\Res}{Res\s{}  }

\newcommand{\bma} {\begin{pmatrix}}
\newcommand{\ema} {\end{pmatrix}}

\newtheorem*{lemma}{Lemma}

\newcommand{\iu}{{i\mkern1mu}}
\newcommand{\Vl}{V^{\text{L}}}
\renewcommand{\d}[1]{\ensuremath{\operatorname{d}\!{#1}}}
\newcommand{\me}{\mathrm{e}}
\newtheoremstyle{dotless}{}{}{\itshape}{}{\bfseries}{}{ }{}
\theoremstyle{dotless}

\newcommand{\qm} {\varrho}

\newcommand{\s}{\hspace{.08em}}
\newenvironment{sloppypar*}{\sloppy\ignorespaces}{\par}

\let\oldFootnote\footnote
\newcommand\nextToken\relax
\renewcommand\footnote[1]{\oldFootnote{#1}\futurelet\nextToken\isFootnote}
\newcommand\isFootnote{\ifx\footnote\nextToken\textsuperscript{,}\fi}

\usepackage[nottoc]{tocbibind}

\usepackage{authblk}
\makeatother
\author[1]{Michael Kreshchuk}
\author[1,2]{Tobias Gulden}
\affil[1]{
{\small~ School of Physics and Astronomy, University of Minnesota, Minneapolis MN 55455 USA}
}
\affil[2]{
{\small~Technion~---~Israel Institute of Technology, 3200003, Haifa, Israel}
}
\affil[ ]{
{\small{e-mails: ~\href{mailto:kreshchu@physics.umn.edu}{kreshchu @ physics.umn.edu}
   ~,~~~~  \href{mailto:gulden@physics.technion.ac.il}{gulden @ physics.technion.ac.il}
}}}
\title{
{\Large{\textbf{
A duality of the classical action yields a reflection symmetry of the quantum energy spectrum
}}}}
\date{}
\begin{document}
\maketitle
\begin{abstract}



The concept of duality reflects a link between two seemingly different physical objects. An example in quantum mechanics is a situation where the spectra (or their parts) of two Hamiltonians go into each other under a certain transformation. We term this phenomenon as {\it{the energy-spectrum reflection symmetry}}.

We develop an approach to this class of problems, based on the global properties of the Riemann surface of the quantum momentum function, a natural quantum-mechanical analogue to the classical momentum. In contrast to the algebraic method, which we also briefly review, our treatment provides an explanation to the long-noticed matching of the perturbative and WKB expansions of dual energy levels. Our technique also reveals the classical origins of duality.

\end{abstract}
\newpage
\tableofcontents
\newpage
\section*{Preamble\label{preambuloid}}
\addcontentsline{toc}{section}{\nameref{preambuloid}}

The concept of \textit{duality} is widely used in modern physics~\cite{Dual,Mald,tHooft,Sussk,mirr,Seib}. Establishing of connections between different regimes of a system (or between different systems) often yields novel results, such as non-perturbative solutions or hidden symmetries~\cite{leeyi,wyi,monod,igna,ghost,shata}. Such results are usually achieved through analysing the properties of the underlying mathematical structures~---~algebras, manifolds, etc.

The term \textit{duality}, in the meaning we use it here, was introduced in~\cite{Dual} in the studies of the Lam\'e system. There,  potentials with different values of the elliptic parameter $\s \nu\s $ were named dual if their spectra (i.e., the band-gap edges) interchange under the switch $\s E\s \to\s -\s E\s $. In the \textit{self-dual} case (the one corresponding to $\s \nu=1/2\s $), the spectrum of the potential maps onto itself under this switch. One way to understand this phenomenon comes from  algebraisation of the Lam\'e potential, i.e., through writing the Hamiltonian in a matrix form.

The property of the spectrum to transform into itself under the change $\s E \to -E\s $ was originally established in the studies of a sextic quasi-exactly solvable (QES) potential~\cite{Shifman,ShifmanTurbiner}. This property was termed as $\s ${\it{energy reflection (ER) symmetry}}$\s $. $\s $In our article, we shall refer to it as $\s ${\it{energy spectrum reflection (ESR) property}}$\s $. In contrast to the Lam\'e model, where \textit{all} band-gap edges can be found by matrix methods~\cite{Dual}, the QES problems normally permit only for partial algebraisation of the (generally infinite) spectrum~\cite{turb1988}. Consequently, the ESR symmetry refers only to its algebraic part. In this restricted sense the sextic potential can be regarded as \textit{self-dual}. Therefore, it is natural to hypothesise that there also exist \textit{dual} sextic potentials whose algebraic parts of the spectra interchange under ${\s E\s \to\s -\s E\s }$.

We show that such potentials indeed exist, and prove their duality both by matrix methods and by using the global properties of the Riemann surface of the quantum momentum function. Then, we elaborate in detail on the latter of these two approaches.
This allows us, inter alia, to resolve the little puzzle which was pointed at in~\cite{Shifman} and also discussed in~\cite{Dual}. Namely, the aforementioned \textit{duality} of energy levels implies the precise matching of the perturbative expansion of the low levels with the WKB expansion of the dual high levels~--- which na\"ively seem to have completely different nature. By means of the generalised Bohr-Sommerfeld quantisation condition (see \cite{Bender} or Appendix \ref{miauu} for details) we explain their intimate relationship.

In~\cite{Alvarez} it was spelled out for the first time that the symmetry of the quantum energy levels of a QES system stems from a symmetry of the corresponding classical system. In {\it{Ibid.}}, the symmetry of the classical system was discussed in the canonical language of phase space. In the present work, we shall show how classical and quantum symmetries can be uniformly described in the language of the action, for which, as we shall demonstrate, the term \textit{duality} can be applied as well. This formalism allows one to investigate at a deeper level the connection between the symmetries of the classical and quantum systems. Specifically, it turns out that such symmetries emerging in both classical and quantum cases can be understood through studying the global properties of the Riemann surfaces of the classical and quantum momenta, as was suggested in \cite{HamJa}.

In Sections~\ref{oneone} and~\ref{onetwo} we define the Riemann surfaces of the canonical momentum as a function of position. There we show that the classical sextic and elliptic potentials possess a special property which we term the duality of the actions, equations~\eqref{S} and~\eqref{Ss}. Combined with the Bohr-Sommerfeld quantisation condition, this property allows to readily extend the notion of duality to the quantum-mechanical spectrum at the perturbative level. In Section~\ref{calcuclas} we briefly review some results from the upcoming work \cite{SHORT} where a more detailed analysis of the potentials' Riemann surfaces is presented.

In Section~\ref{twoo} we address two quantum-mechanical problems, the sextic quasi-exactly solvable and the Lam\'e  potentials. We briefly review the algebraic approach to these problems, and then show how the analysis performed in Section~\ref{sectionone} can be extended to quantum mechanics at the non-perturbative level. This allows us to explain the matching of the perturbative and WKB expansions for the dual levels to any order in the coupling constant.

\section{Duality, self-duality, and action reflection symmetry in classical mechanics \label{sectionone}}

In this section we derive the duality property of the classical action for the sextic and elliptic potentials, see equations~\eqref{S} and~\eqref{Ss} below. These derivations lay the basis for the subsequent results on the symmetry between the quantised counterparts of those systems. We also discuss how the Picard-Fuchs equation can be used for finding the classical action.


\subsection{Sextic potential\label{oneone}}
\begin{sloppypar}
In~\cite{Shifman,ShifmanTurbiner}, an interesting property of the sextic quasi-exactly solvable potential has been first observed. A part of its spectrum turned out to be symmetric with respect to the ${\s E \to-E\s}$ transformation. Moreover, it was observed that this statement was valid order by order in the perturbation theory. The latter fact is not as obvious as it may seem, since the said matching only takes place for a discrete set of values of the coupling constant. The traditional algebraic approach to the quasi-exactly solvable problems provides no clue to explaining this mystery. The puzzle can, however, be resolved by looking at the problem from a different perspective. As will be seen shortly, the aforementioned symmetry of a part of the spectrum stems from a duality property of the corresponding classical action.
\end{sloppypar}

We begin with exploring the one-dimensional motion in the bounded potential
\begin{equation}
\begin{gathered}
    V_{a,\s c}(x) = a + b\s x^2 + c\s x^4 + d \s x^6\quad,\\
    a,\s b,\s c,\in \mathbb{R}\quad,\qquad d\in\mathbb{R}^+\quad.
    \label{Vcl}
\end{gathered}
\end{equation}

Our study will address the abbreviated classical action \footnote{~The so-defined action is but a Legendre transform of the standard action~$\int^t L(q, \dot{q})\s \d t\s $, with $L$ being the Lagrangian.}
\begin{equation}
    S(E) = \oint \limits_{\mathcal{C}_R} p(x, E)  \d x \quad,
    \label{18}
\end{equation}
where $p(x, E) = \sqrt{2\s (E - V(x))}$ is the canonical momentum with the mass set to be $m=1$. The contour $\mathcal{C}_R$ encloses two real turning points $x_1$ and $x_2$ between which $V(x) < E\,$.\,\footnote{~It will be agreed that $\,x_1<x_2\,$, so that  $V(x) < E$ holds for $x_1 < x < x_2$.
Also, the contour $\mathcal{C}_R$ should be close enough to the real axis, so that it does not enclose other branch points or singularities.}
A pair of two sextic potentials with opposite values of the parameters $a$ and $c$ possesses the symmetry property
\begin{equation}\label{potsymm}
    V_{-a,\s -c}(x) = - V_{a,\s c}(\iu\s x) \quad.
\end{equation}
This is a generalisation of the case considered in~\cite{Shifman,ShifmanTurbiner} which addressed a sextic potential with vanishing $a=c=0$ and the resulting symmetry $V_{0,0}(x)=-V_{0,0}(\iu x)$. From the symmetry property~\eqref{potsymm}, we will derive a duality property for the classical action \eqref{18}.
In the light of this property, the special case considered in~\cite{Shifman,ShifmanTurbiner} will be naturally referred to as {\it{self-duality}}.

In~\eqref{18}, the integral along the closed contour $\mathcal{C}_R$ is nonzero due to the presence of a branch cut of the momentum inside it, between the turning points.$\s $\footnote{~Generally speaking, one has freedom in defining the branch cuts. For the real turning points, it is usually convenient to make the cut go along the finite interval of the real axis between the turning points.}
This makes the contour non-contractible or, in the parlance of topology, homotopically non-trivial.

The momentum is set positive along the bottom of the cut, in order for the action to stay positive when the direction of the contour in~\eqref{gammas} is chosen  counter-clockwise.

\begin{figure}[!ht]
\centering
\subfloat[][
The potential $V_{a,\s c}(x)$ for $b<0$.
]{
\includegraphics[width=.45\textwidth]{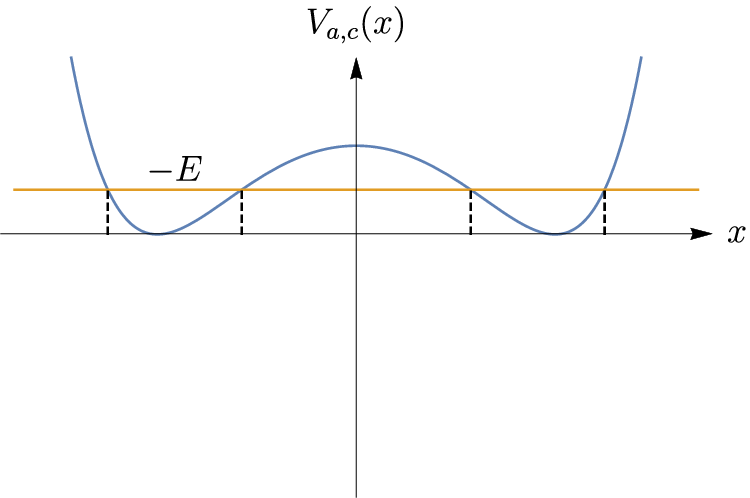}
\label{aug11}
}\qquad
\subfloat[][
The potential $V_{-a,-c}(x)$ for $b<0$.
]{
\centering
\includegraphics[width=.45\textwidth]{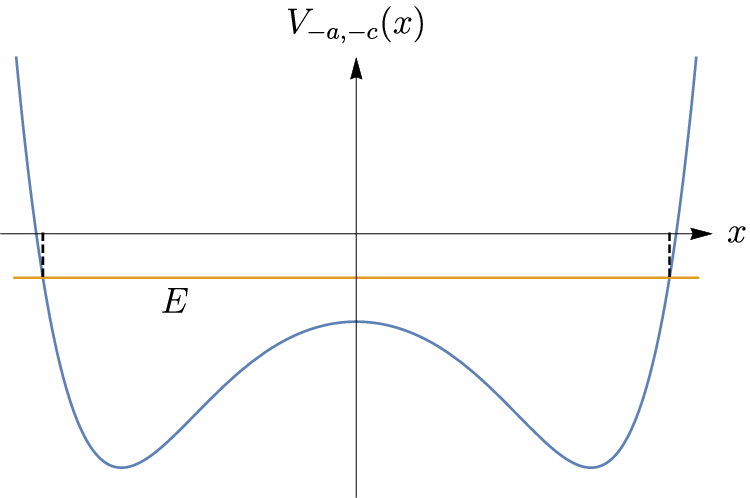}
\label{aug12}
}
\\
\subfloat[][
The potential $V_{a,\s c}(x)$ for $b>0$.
]{
\includegraphics[width=.45\textwidth]{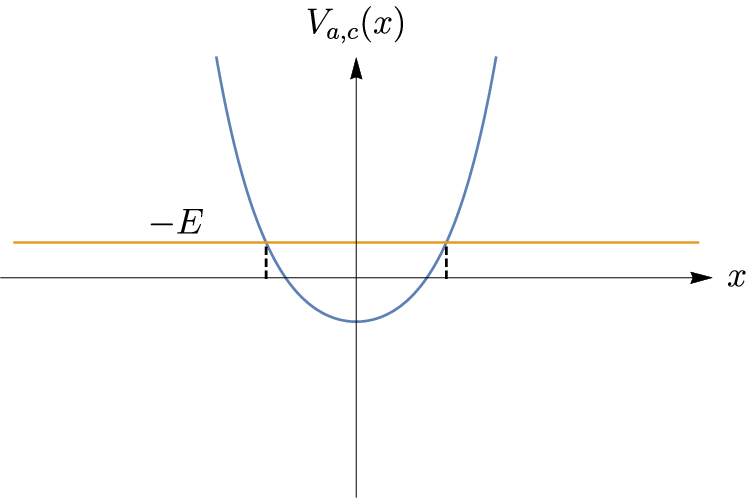}
}\qquad
\subfloat[][
The potential $V_{-a,-c}(x)$ for $b>0$.
]{
\includegraphics[width=.45\textwidth]{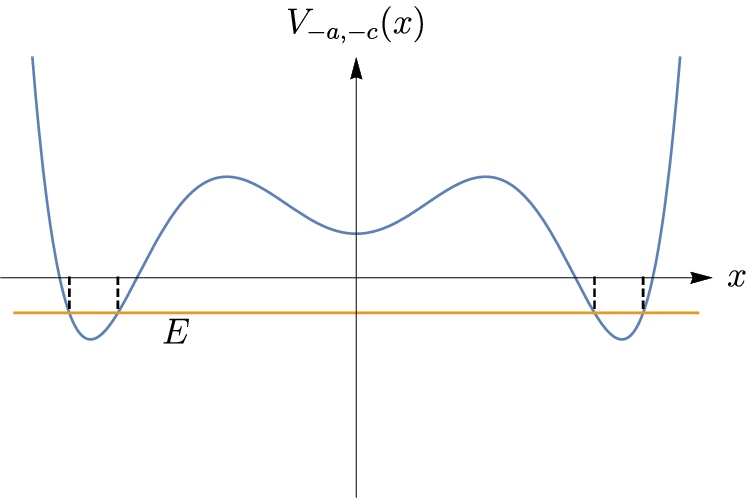}
}
\caption{}
\label{F1}
\end{figure}

The classical turning points are the real roots of the equation
\begin{equation}
    p_{a,\s c}(x, E) = 0.
\end{equation}
For various values of the parameters entering the potential, and for various values of energy, it may have up to six real roots. This is illustrated by Figure~\ref{F1} which also shows how the sign of $b$ changes the overall shape of the potential.

For $b<\dfrac{c^2}{4d}\s$, there exists a range of energies for which one of the equations
\begin{subequations}
\begin{alignat}{9}
    &V_{a,\s c}(x)  &&=-&&E \quad,\label{9a}\\
    &V_{-a,-c} (x) &&=&&E \quad,\label{9b}
\end{alignat}
\end{subequations}
has four real solutions, while another equation has only two. Without loss of generality we assume below that the energy $E$ stays in the range where~\eqref{9b} has two solutions, while~\eqref{9a} has four.\footnote{~See Appendix~\ref{shapes} for a detailed discussion.
}

\begin{figure}
\centering
\begin{minipage}{0.45\textwidth}
\centering
\includegraphics[width=1\textwidth]{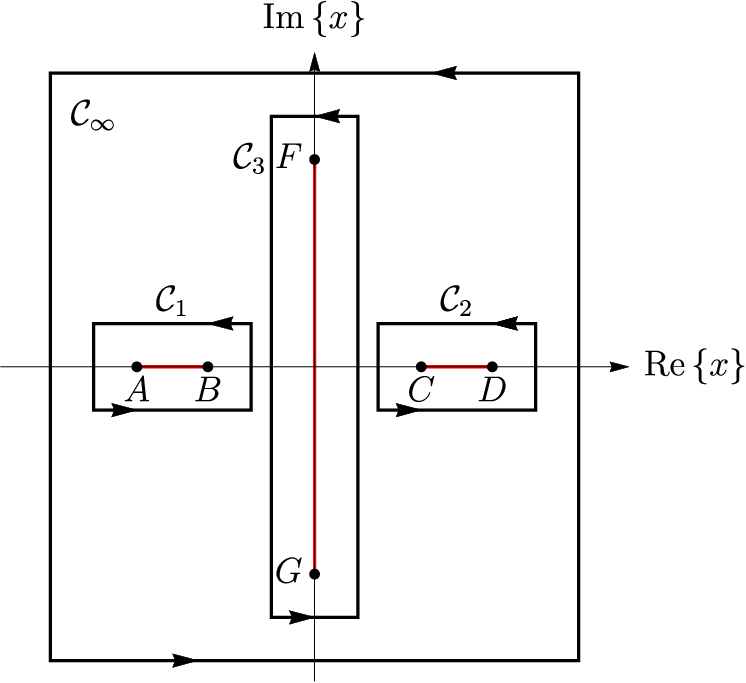}
\caption{The integration cycles\\for $p_{a,\s c}(x,\s -E)$.}\label{roots6}
\end{minipage}\hfill
\begin{minipage}{0.45\textwidth}
\centering
\includegraphics[width=1\textwidth]{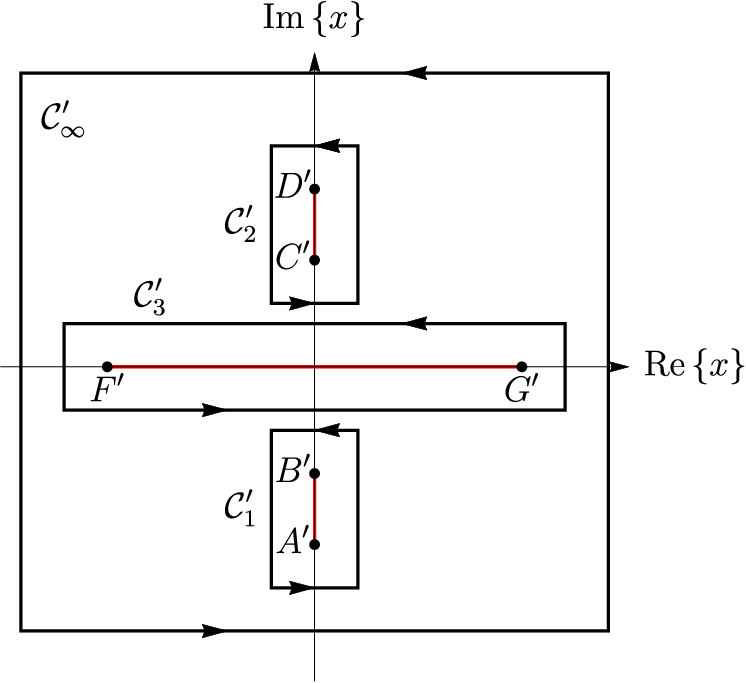}
\caption{The integration cycles\\for $p_{-a,-c}(x, E)$.}\label{roots62}
\end{minipage}
\end{figure}

Let us write down the action for (a) the potential $\s V_{a,\s c}(x)\s $ and energy $(-E)$, and (b) the potential $\s V_{-a,-c}(x)\s $ and energy $E$. We define the \textit{periods} $\s \gamma_{n}(E)\s $ and $\s \gamma_{n}^{\s \prime}(E)\s $ as
\begin{equation}
\begin{array}{cc}
{\displaystyle
\gamma_{n}(E)=~\oint \limits_{\mathcal{C}_n} p_{a,\s c}(x,\s -E)\d x} \\
~\\
{\displaystyle
\gamma_{n}^{\s \prime}(E)=~\oint \limits_{\mathcal{C}_n'} p_{-a,-c}(x, E)\d x}
\end{array}
\qquad\qquad,\qquad
n=1,2,3 \quad,
\label{gammas}
\end{equation}
with $\s {\mathcal{C}_n}\s $ and $\s {\mathcal{C}^{\s \prime}_n}\s $ being the cycles in the complex $\s x$-planes of $\s p_{a,\s c}(x,\s -E)\s $ and $\s p_{-a,-c}(x, E)\s$ (see figures \ref{roots6} and \ref{roots62}).\footnote{~Be mindful that Figure~\ref{roots6} represents only one of the two sheets of the Riemann surface of the momentum. The second sheet corresponds to the negative sign in front of the square root.
However, it turns out that all the calculations needed in this section can be performed on a single sheet.}

\begin{sloppypar}
Now we shall express the action of the potential $V_{-a,-c}(x)$ in terms of the periods of the potential $V_{a,\s c}(x)$.
Since ${\,p_{a,\s c}(x,\s  E) = \iu\s p_{-a,\s -c}(\iu\s x,-E)\,}$, then Figures~\ref{roots6} and~\ref{roots62} are identical up to a rotation by an angle of $\pi/2$. Consequently, identical are
the integrals over the contours $\s {\mathcal{C}_n}\s $ and $\s {\mathcal{C}_n^{\s \prime}}\s $. Indeed,
\end{sloppypar}
\begin{eqnarray}
\begin{gathered}
    \gamma_1 (E) = \gamma_2 (E) =
    \oint \limits_{\mathcal{C}_1} p_{a,\s c}  (x, \s -E) \d x =
    2  \int  \limits_A^B p_{a,\s c}  (x, \s -E) \d x =
    2  \int  \limits_A^B p_{-a,-c}  (\iu\s x, \s E) \d (\iu\s x) \\=
    2  \int  \limits_{A'}^{B'} p_{-a,-c}  (x, \s E) \d x =
    \oint \limits_{\mathcal{C}_1'} p_{-a,-c}  (x, \s E) \d x =
    \gamma_1^{\s \prime} (E) = \gamma_2^{\s \prime} (E) \quad.\label{rotate}
\end{gathered}
\end{eqnarray}
A similar derivation renders $\gamma_3 (E)= \gamma_3^{\s\prime} (E)$.
The periods are linked to the actions as
\begin{subequations}\label{scy}
\begin{alignat}{9}
    S_{a,\s c} (-E)
    &\equiv \gamma_{1,2} (E) &&= \gamma_{1,2}^{\s \prime} (E)\quad&&,\\
    S_{-a,-c} (E)
    &\equiv \gamma_{3}^{\s \prime} (E) &&= \gamma_{3} (E)\quad&&.
 \end{alignat}
 \end{subequations}
Hereafter primes will be omitted, because the equalities~\eqref{scy} make them redundant.

It will be helpful to introduce the period
\begin{equation}
    \gamma_{\infty}(E) \equiv
    \oint \limits_{\mathcal{C}_{\infty}} p_{a,\s c}(x,\s -E)\d x =
    \gamma_{1}(E) + \gamma_{2}(E) + \gamma_{3}(E) \quad.
    \label{gammas6}
\end{equation}
This is an integral over a contour $\mathcal{C_\infty}=\mathcal{C}_1+\mathcal{C}_2+\mathcal{C}_3$ enclosing all the branching points, which is equivalent to a countour enclosing the point at infinity. So evaluation of this integral can be carried out with the aid of the residue at infinity:
\begin{equation}
    \gamma_\infty (E) =
     -2 \pi \iu \Res \{ p_{a,\s c}(x, -E)dx, \infty \}
    = \dfrac{\pi  \left(c^2-4\s b\s d\right)}{(2\s d)^{3/2}}\quad.
    \label{eq:gammainf}
\end{equation}
The key property of this result, which turns out to be crucial for the entire subsequent analysis, is that $\gamma_{\infty}$ does \textit{not} depend on the energy.
In other words, $\gamma_{1,2}(E)$ and $\gamma_3(E)$, as functions of energy, turn out to be linearly dependent. A similar result can be obtained for $V_{-a,-c}(E)$. Thus we arrive at
\begin{equation}
    2S_{a,\s c}(-E) + S_{-a,-c} (E)
    = 2S_{-a,-c}(-E) + S_{a,\s c} (E)
    = \dfrac{\pi  \left(c^2-4\s b\s d\right)}{(2\s d)^{3/2}}\quad, \label{S}
\end{equation}
an equality to be referred to as \textit{duality of the action}.\s\footnote{
For the sum $~2S_{\pm a,\pm c}(-E) + S_{\mp a,\mp c} (E)~$
to come out energy-independent, not only the duality property was essential but also a simple analytic structure of the potential. Had we started with potentials of a higher order, we would have encountered more branching points in the complex plane. So the sum of the periods responsible for the classical actions would no longer be proportional to the residue at infinity, because other nontrivial periods
(enclosing additional branching points)
would have emerged:
\begin{subequations}
\begin{equation}
    \gamma_\infty
    = \gamma_1(E) + \gamma_2(E) + \gamma_3(E) + \gamma_4(E) +\ldots \quad.
 \label{26a}
\end{equation}
The latter is equivalent to
\begin{equation}
\gamma_1(E) + \gamma_2(E) + \gamma_3(E)=\gamma_\infty-\gamma_4(E)-\ldots \quad,
\label{26b}
\end{equation}
\label{26}
\end{subequations}
with the right-hand side (and, therefore, also the left-hand side) being energy-dependent.
}

From the classical point of view, equation~\eqref{S} is merely an interesting mathematical observation.
It, however, acquires far-going physical consequences at the quantum-mechanical level. These become apparent if we recall that the action as a function of  energy shows up in the Bohr-Sommerfeld quantisation condition from which the energy levels of the quantum system can be derived to the first order in $\hbar$.
However, in what follows we will go even further: by generalising the classical action to its natural quantum counterpart, we will be able to make \textit{exact} statements about the energy levels of the two systems. Consequently, both  equation~\eqref{S} and its quantum analogue, equation \eqref{GBSa} below, will serve as  conditions interrelating the spectra of the two systems~--- i.e., will embody
the \textit{energy spectrum reflection} property.

\subsection{Elliptic potential \label{onetwo}}
We now derive a similar duality relation as in equation~\eqref{S} for the action of the elliptic potential
\begin{equation}\begin{gathered}
V(x|\nu) = a\s\nu \s\snt (x| \nu) + b \quad,\\
a \in \mathbb{R}^+\quad,\qquad b \in \mathbb{R} \quad,\qquad\nu \in [0,1]\quad,
\label{Ve}
\end{gathered}\end{equation}
where $\sn (x, \nu)$ is the Jacobi elliptic function~\cite{NIST}.
This is the classical version of the Lam\'e potential that was intensely studied  in terms of duality and energy spectrum reflection symmetry~\cite{Dual,Kha,shata}.

\begin{figure}
\centering
\begin{minipage}{0.48\textwidth}
\centering
\includegraphics[width=1\textwidth]{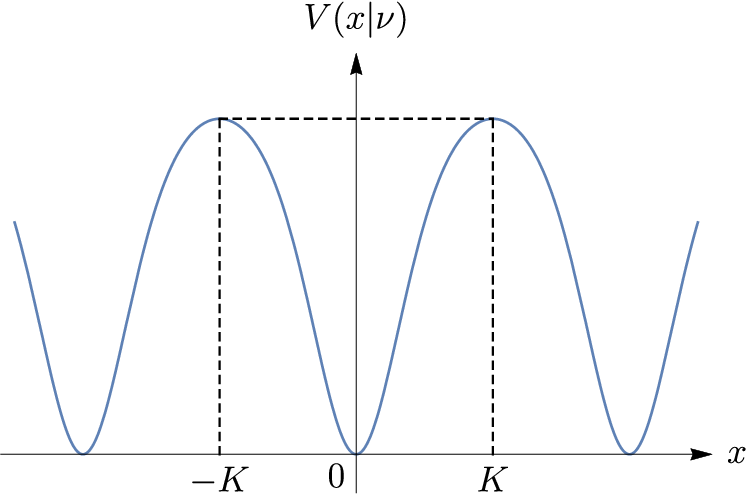}
\caption{The periodic potential $V(x|\nu)$ along the real line of $x$.}\label{vxnu}
\end{minipage}\hfill
\begin{minipage}{0.48\textwidth}
\centering
\includegraphics[width=1\textwidth]{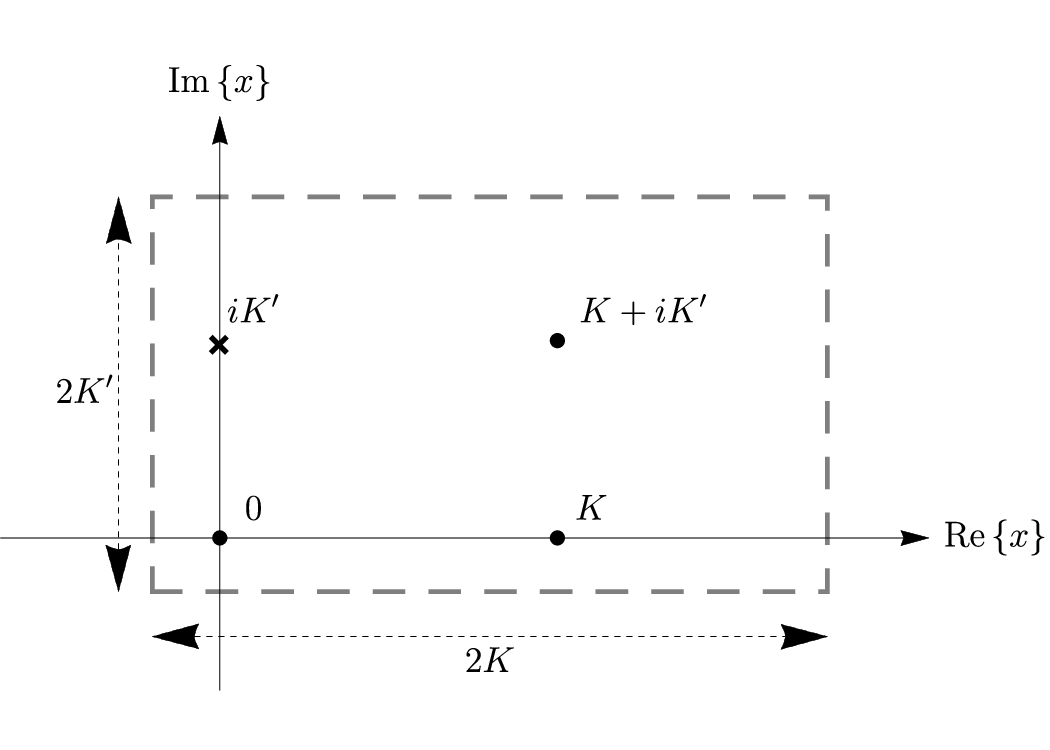}
\caption{The fundamental parallelogram of the potential  $V(x|\nu)$ with a pole at $x=\iu K'$.}\label{vxnuriem}
\end{minipage}
\end{figure}

The potential~\eqref{Ve} is a doubly periodic meromorphic function. Its real and imaginary periods are ${2\s K \equiv 2\s K(\nu)}$ and ${2\s \iu\s  K' \equiv 2\s \iu\s  K(1-\nu)}$, where
\begin{equation}
    K(\nu) = \int \limits_0^{\pi/2}
    \dfrac{\d\theta}{\sqrt{1-\nu\s \sin^2 \theta}}
\end{equation}
is the complete elliptic integral of the first kind. After gluing the opposite sides of the fundamental parallelogram (Figure~\ref{vxnuriem}), one obtains a torus (deprived of a point corresponding to the pole).

The potential $V(x|\nu)$ has a second-order pole at $\iu K'$ and a second-order zero at $x=0$. Hence there exist two turning points:
\begin{equation}
    V(x|\nu) = E \qquad\Longrightarrow\qquad x_1,\s x_2\quad
\end{equation}
which are real iff the energy assumes values between the minimum and the maximum of the potential.

The potential $V(x\s |\s 1-\nu)$ is related to the one defined in~\eqref{Ve} by symmetry, in the sense that
\begin{equation}
V(\iu\s x + K + \iu\s K'\s|\s\nu) + V(x\s|\s1 - \nu) = a+2\s b \quad.
\end{equation}
For the maxima of these potentials, it will be convenient to introduce the notations
\begin{equation}\begin{aligned}
    V_{\nu,\s\text{max}} = V(K\s|\s\nu) = a\s\nu + b \quad,\qquad
    V_{1-\nu,\s\text{max}} = V(K'\s|\s1-\nu) = a\s(1-\nu) + b \quad.
\end{aligned}\end{equation}

For the energies above the maximum of the potential, the motion in the potential~\eqref{Ve} is qualitatively distinct from that in the polynomial potential: the motion is unbounded but still periodic.\s\footnote{~When a particle moving in the periodic potential reaches an end of the interval, it immediately appears on the opposite end, not bounces back.}

We now turn to the analysis of the Riemann surface of the canonical momentum
\begin{equation}
p\s(x,E\s|\s\nu) = \sqrt{E  - a\s\nu\s\snt(x,\s\nu)-b}\quad,
\label{plame}
\end{equation}
where we set the mass to be $\s m=1/2\s $. The symmetry property for the momentum looks as
\begin{equation}\begin{aligned}
p\s(\iu\s x+K+\iu\s K',\s V_{\nu,\s\text{max}}-\epsilon\s|\s\nu) =
\iu\s p\s(x, V_{1-\nu,\s\text{max}}+\epsilon\s|\s1-\nu) \quad.
\end{aligned}
\end{equation}

\begin{figure}
\centering
\includegraphics[width=1\textwidth]{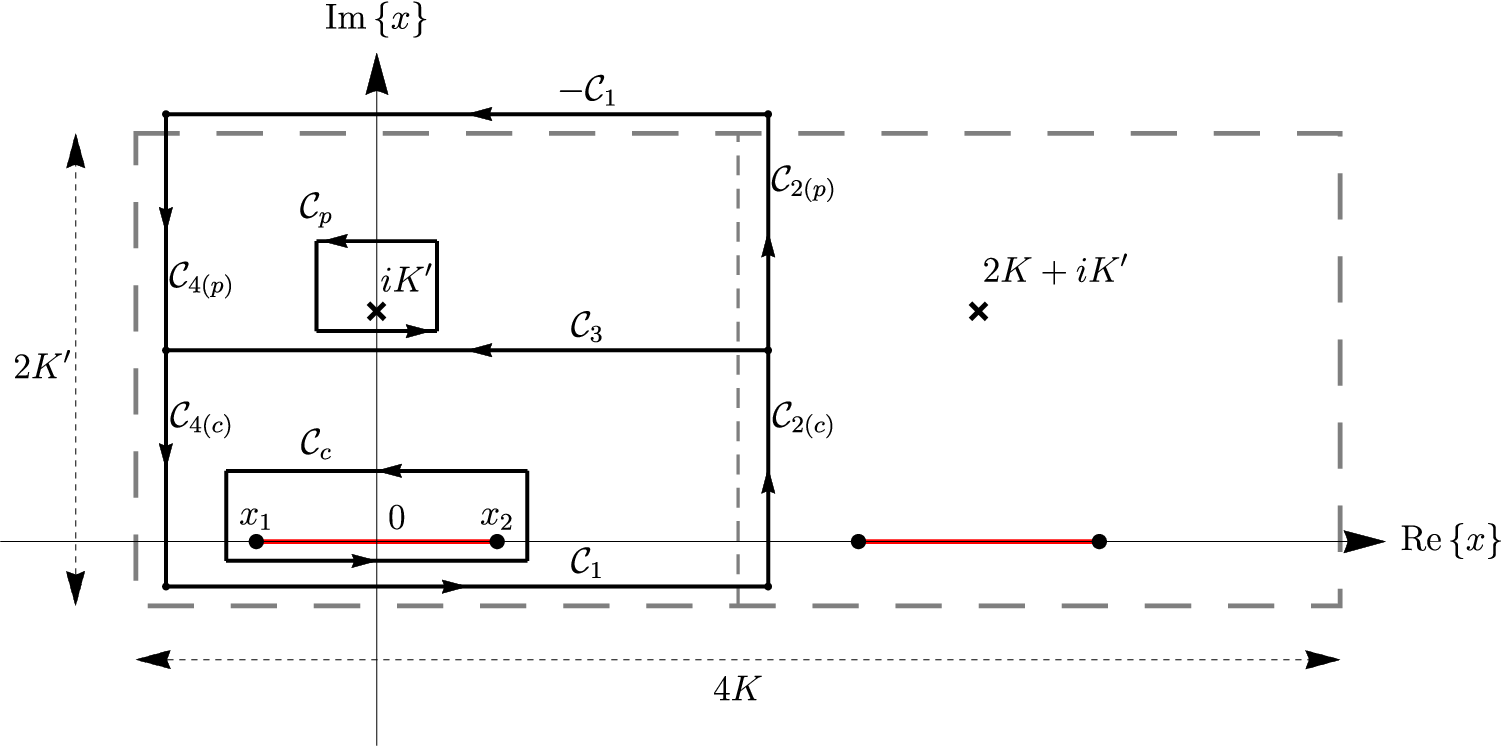}
\caption{The integration contours on the Riemann surface of the $p\s (x, V_{\nu,\s\text{max}}-\epsilon\s |\s \nu)$.}
\label{pxnuriem}
\end{figure}

The double periodicity of potential~\eqref{Ve} carries over to momentum~\eqref{plame}. However, owing to the presence of the square root, the period in the real direction is doubled and equals $4\s K$. Thus, we can treat the momentum $p\s(x, E|\s\nu)$ as a doubly periodic function whose real and imaginary periods are $4\s K$ and $2\s\iu\s K'$, correspondingly. The two halves of the fundamental parallelogram in Figure~\ref{pxnuriem} correspond to the two sheets of the square root. On each sheet, the function has a cut and a pole of the first order. One can travel between the sheets either by moving parallel to the real axis or by crossing the branch cuts. After gluing the opposite sides of the parallelogram and the cuts, one ends up with a genus-$2$ Riemann surface deprived of two points.

We introduce the integration contours for the momentum $p\s (x, V_{\nu,\s\text{max}}-\epsilon\s |\s \nu)$ as shown in Figure~\ref{pxnuriem}:
\begin{equation}\begin{alignedat}{4}
    \mathcal{C}_c &\equiv &&\mathcal{C}_1 + \mathcal{C}_{2(c)} + \mathcal{C}_3 + \mathcal{C}_{4(c)} \quad,\qquad
    \mathcal{C}_{2} &&=  \mathcal{C}_{2(c)} +  \mathcal{C}_{2(p)} \quad, \\
    \mathcal{C}_p &\equiv -&&\mathcal{C}_3 + \mathcal{C}_{2(p)} - \mathcal{C}_1 + \mathcal{C}_{4(p)} \quad,\qquad
    \mathcal{C}_{4} &&=  \mathcal{C}_{4(c)} +  \mathcal{C}_{4(p)} \quad.
    \label{Ccp}
\end{alignedat}\end{equation}
The contours $\mathcal{C}_c$, $\mathcal{C}_p$, $\mathcal{C}_2$ and $\mathcal{C}_4$ are closed and homotopically non-trivial. However we want to note that $\mathcal{C}_1$ and $\mathcal{C}_3$ are not closed, continuing them along the second sheet gives closed contours. What precludes us from writing $(- \mathcal{C}_2)$ instead of $\mathcal{C}_4\;$ (the way we did it with $\mathcal{C}_1$ in the definition of $\mathcal{C}_p$ in Figure~\ref{pxnuriem}) is the presence of the second sheet of the Riemann surface with singularities on it. In other words, we cannot continuously deform $\mathcal{C}_4$ into $(-\mathcal{C}_2)$.

After summing up the two lines of~\eqref{Ccp}, one arrives at
\begin{equation}\label{Lamecontours}
    \mathcal{C}_c + \mathcal{C}_p
    = \mathcal{C}_2 + \mathcal{C}_4 \quad.
\end{equation}
Let us now define the following integrals along these contours:
\begin{equation}\begin{alignedat}{2}
    \gamma_c\s (\epsilon) &= \phantom{-}\oint \limits_{\mathcal{C}_c} p\s (x, V_{\nu,\s\text{max}}-\epsilon\s |\s \nu) \s \d x \quad,\\
    \gamma_n\s (\epsilon) &= - \oint \limits_{\mathcal{C}_n} p\s (x, V_{\nu,\s\text{max}}-\epsilon\s |\s \nu) \s \d x
    \end{alignedat}
    \quad,\qquad
    n=p,\s 2,\s 4\quad.
    \label{ellcont}
\end{equation}
The signs are chosen for the integration periods to stay positive when the direction along the contours is set counter-clockwise and the  momentum is set positive along the bottom of the cut. According to equation \eqref{Lamecontours}, these integration periods obey an equality similar to~\eqref{gammas6}:
\begin{equation}
    \gamma_{c}\s (\epsilon) + \gamma_2\s (\epsilon) + \gamma_4\s (\epsilon) = \gamma_{p}\s (\epsilon)\quad.
    \label{29}
\end{equation}
The integral along the contour enclosing the pole can be evaluated with the aid of the residue:
\begin{equation}
    \gamma_p\s (\epsilon)
    = -2\s \pi\s \iu\s \Res \{ p\s (x, V_{\nu,\s\text{max}}-\epsilon\s |\s \nu)dx,\s  \iu\s K' \}
    = -2\s \pi\s \iu (\iu\s \sqrt{a})
    = 2\s \pi\s  \sqrt{a}\quad.
    \label{Ssres}
\end{equation}
Like in equation \eqref{eq:gammainf}, this residue is independent of energy, which is crucial for the duality of the action.
Now is the right time to identify the integration periods with the appropriate actions:
\begin{subequations}\begin{alignat}{4}
    S\s (V_{\nu,\s\text{max}}-\epsilon\s |\s \nu)
    &\s \equiv \s &\gamma_c \s (\epsilon)& \quad,\\
    S\s (V_{1-\nu,\s\text{max}} + \epsilon\s |\s 1-\nu)
    &\s \equiv &\gamma_2 \s (\epsilon)\s &
    =\gamma_4 \s (\epsilon) \quad.
\end{alignat}
\label{31}
\end{subequations}
Together, the formul\ae~\eqref{29},~\eqref{Ssres} and~\eqref{31} entail the duality relation between the actions of the two  potentials:
\begin{equation}
    S(V_{\nu,\s\text{max}}-\epsilon\s |\s \nu) + 2\s S(V_{1-\nu,\s\text{max}}+\epsilon\s |\s 1-\nu)
    = 2\s S(V_{\nu,\s\text{max}}+\epsilon\s |\s \nu) + S(V_{\nu,\s\text{max}}-\epsilon\s |\s 1-\nu)
    = 2\s \pi\s \sqrt{a}\quad,
    \label{Ss}
\end{equation}
which is a close analogue to the relation~\eqref{S}. As in the case of the sextic potential, at the classical level this duality relation is merely a curious mathematical fact. However after quantisation it leads to the energy spectrum reflection property of the Lam\'e potential that was studied in \cite{Dual,Kha,shata}.

It is worth noting that the origin of the factor of 2 in equation~\eqref{S} is slightly different from that of the one emerging in~\eqref{Ss}. In the former case, one has two symmetric wells in the potential which merge above the local maximum. In the latter, above the maximum we have two options for the particle~---~either to move from left to right or from right to left. In both cases, the action is an integral along a non-contractible loop.

\subsection{Picard-Fuchs Equation and Classical Action \label{calcuclas}}

This section briefly summarises the results of the upcoming publication~\cite{SHORT}. There we demonstrate that the classical action for the two afore-discussed classical systems can be found with the aid of the Picard-Fuchs equation. This equation is a powerful tool from algebraic topology that was introduced in high-energy physics through the Seiberg-Witten solutions of supersymmetric models~\cite{SeibergWitten1,SeibergWitten2} and was subsequently applied to integrable systems~\cite{DonagiWitten,Gorsky} and many other problems.

Apart from its mathematical value and its applicability to classical mechanics, in \cite{SHORT} we also show how this kind of analysis can be extended to the quantum-mechanical level. This is so owing to the fact that the quantum corrections to the classical action inhabit the same Riemann surface as the classical momentum itself.

\subsubsection{{Self-dual sextic potential \label{sdsp}}}

For the ease of notation, we shall focus on the self-dual case of the sextic potential which is given by the potential $V_{0,0}(x)=b\s x^2+d\s x^6$ from equation \eqref{Vcl}, and shall assume $b<0$. The momentum
has the explicit form
\begin{equation}
    p(x, E) = \sqrt{2(E-b\s x^2-d\s x^6)},
\end{equation}
while the classical action \eqref{18} becomes
\begin{equation}
    S(E) = \oint \limits_{\mathcal{C}_R}p(x, E)\d x = \dfrac{\sqrt{2}(-b)^{3/4}}{d^{\s 1/4}}\oint \limits_{\mathcal{C}_R}\sqrt{\dfrac{2}{3^{3/2}}u+y^2-y^6}\d y = \dfrac{\sqrt{2}(-b)^{3/4}}{d^{\s 1/4}}\s  \tilde{S}(u)\quad.
    \label{Ssexticdual}
\end{equation}
Here we rescaled the coordinate as $y=(-d/b)^{1/4}x$ and the energy as $u=\dfrac{3\sqrt{3\s d}}{2(-b)^{3/2}}E$. Then $\tilde{S}(u)$ is the action of a sextic double-well potential similar to Figure~\ref{F1}b where the minimum of the potential lies at $u=-1$ and the central maximum is at $u=0$.

The momentum is a globally double-valued function. Therefore, its Riemann surface consists of two sheets, each having six branch points (the roots of $p(x,u)=0$) and, consequently, three branch cuts. Connecting the sheets along the cuts gives birth to a genus-2 surface. Additionally, there is one singiarity on each sheet, the pole at $x=\infty$. On this surface, one can draw up to five homotopically distinct closed cycles, i.e., cycles none of which can be continuously deformed into another. By an algebraic-topology argument, we can deduce that the number of linearly independent 1-forms on this surface is equal to the number of homotopically distinct cycles.\s\footnote{~In a  more accurate mathematical language, our statement sounds like this: {\it{the first homology and first cohomology groups are isomorphic for the surface considered}}. Here one may be tempted to refer to the Poincar\'e duality, i.e., to the isomorphism between the {$k$-th} cohomology group and the {$(n-k)$-th} homology groups of an $n$-dimensional manifold. This property, however, does not hold for \textit{punctured} manifolds. Fortunately, a weaker statement, the isomorphism between the first homology and cohomology groups, holds~---~and is sufficient for our needs.

The above statement about isomorphism of the first homology and first cohomology groups is a corollary of the de Rham theorem.} Thus, any six (or more) 1-forms are linearly dependent. For example, consider the set of 1-forms, $\,\{(\partial_u)^n p(x,u)dx\}_{n=0}^5\,$, which comprises the momentum and its first five derivatives with respect to the energy $u$. They cannot be linearly independent. In other words, there must exist a vanishing linear combination of these 1-forms. Then, integration of this combination gives birth to a differential equation for the action:
\begin{equation}
    5 \tilde{S}^{\s \prime}(u) + 59\s u\s \tilde{S}^{\s \prime\prime}(u) + 18 ( 3 \s u^2-1) \tilde{S}^{(3)}(u) + 9 \s u (u^2 - 1) \tilde{S}^{(4)}(u) = 0
    \label{PFsextic}
\end{equation}
This result is known as the \textit{Picard-Fuchs equation}. Its basis solutions can be expressed through the hypergeometric functions $_pF_q$. The classical action is a linear combination of these basis solutions. For a complete derivation and calculation of quantum energy levels we refer the reader to \cite{SHORT}. In \textit{Ibid}., we also provide more mathematical background and discuss why the differential equation \eqref{PFsextic} is of lower order than five, as was initially expected.

\subsubsection{{Elliptic potential \label{lamepfcl}}}

In the case of the elliptic potential \eqref{Ve} the construction of the Riemann surface for the momentum in equation \eqref{plame} is slightly different due to the double-periodicity of the function. Nevertheless, we end up, again, with a genus-$2$ surface with one singularity on each sheet. We introduce $\tilde{S}(u|\nu)$ as:
\begin{equation}
    \label{quiffy}
    S(E|\nu) = \oint \limits_{\mathcal{C}_R}\sqrt{E  - a\s\nu\s\snt(x,\s\nu)-b} \d x = \sqrt{a \nu} \oint\limits_{\mathcal{C}_R} \sqrt{u + \cnt(x,\s\nu)} \d x= \sqrt{a \nu} \s \tilde{S} (u |\nu ) \quad,
\end{equation}
where $u= \dfrac{E-b}{a\s\nu}-1$ so that the minimum of the potential is at $u=-1$ and the maximum at $u=0$. The action is determined by the following Picard-Fuchs equation:
\begin{equation}\begin{alignedat}{9}
    \left(3\s\nu\s u+ 2\s\nu - 1\right) \tilde{S}^{\s \prime}(u|\nu)
    &+
    4\left(\nu\s u(3\s u+4)+\nu-2\s u-1\right) \tilde{S}^{\s \prime\prime}(u|\nu)
    \\&+
    4\s u \left(1+u\right) \left(\nu\s u + \nu - 1\right)  \tilde{S}^{\s \prime\prime\prime}(u|\nu) = 0
    \quad.
\end{alignedat}\end{equation}
Here we show these results for completeness; for a detailed derivation of this equation and a complete analysis of its solutions we refer the reader to \cite{SHORT}.

\section{A symmetry of the quantum energy levels \label{twoo}}

\begin{sloppypar}
In this section, we turn to the main topic of this paper, duality in Quantum Mechanics. Usually, this property is understood as a symmetry of the energy spectrum (or of its part) with respect to the switch ${\s E\to-E\s}$. We, however, will use the word \textit{duality} interchangeably with the way in which we used the term in Section~\ref{sectionone}. While in classical mechanics duality was a symmetry of the abbreviated action, in quantum mechanics it becomes a symmetry of the quantum action functional, and in turn this symmetry entails the reflection symmetry of the energy spectrum.
\end{sloppypar}

After a brief review of the tools to be needed (the most important of those being the generalised Bohr-Sommerfeld quantisation condition \eqref{GBS}\,), we explore in detail the case of the sextic quasi-exactly solvable potential. We begin with the traditional algebraic approach described
by Shifman in~\cite{Shifman}
 where this method was applied to the self-dual cases of the sextic and Lam\'e potentials. We then extend this calculation to the more general case of duality, discussed above in Section \ref{sectionone}.
 Thereafter, we introduce a technique based on the analytic properties of the quantum momentum function, similar to the derivation from Section~\ref{oneone}. \textit{Inter alia}, the latter method allows us to naturally explain the matching of the perturbative and WKB expansions of dual levels.  Further we explore, by similar means, the Lam\'e potential. We conclude this section with an explanation of how the quantum corrections to the classical action can be derived, up to an arbitrary order in $\hbar$, in a manner similar to the way we obtain the Picard-Fuchs equation. The details of this derivation are the central part of an upcoming publication \cite{SHORT}.

Our starting point is the Schr\"odinger equation in one dimension:
\begin{equation}
    \widehat{H} \s \psi(x) = \left[ -\dfrac{\hbar^2}{2\s m} \dfrac{\d{}^2}{\d x^2} + V(x) \right] \psi(x) = E \s \psi (x) \quad.\label{Sch}
\end{equation}
To know the permissible energies, we need to find the spectrum of the differential operator~$\hat{H}$. However the spectra of periodic and non-periodic potentials have to be discussed separately. For bound states in a non-periodic potential the spectrum is discrete. These states are quantum counterparts to periodic bounded motion in a corresponding classical system. This is the only type of motion available for potentials that become infinite at $\s|x|\to\infty\s$. In this case, the process of solving the Schr\"odinger equation gets reduced to diagonalisation of a matrix of a countable (but generally infinite) size. There are few well-known \textit{exactly solvable} problems for which the entire matrix can be diagonalised. In general, though, this kind of problem cannot be solved analytically. However there exist the so-called \textit{quasi-exactly solvable} (QES) problems, the ones permitting partial algebraisation.\footnote{~For a detailed introduction into the topic, see~\cite{Shifman}.} In these problems, the matrix of the Hamiltonian permits to single out a block of a finite size~---~which reduces the search of a limited number of energy levels to the diagonalisation of a finite-size matrix.

For periodic potentials the situation is quite different. Typically, their spectrum is continuous, with gaps in it. The search for the edges of these bands in periodic potentials is analogous, in a certain way, to the search for the bound states in non-periodic potentials. In particular, for the QES periodic potential considered in Section~\ref{ellqes}, it is possible to reduce the problem of finding the locations of the band-edges to diagonalisation of a finite-size matrix.


The spectrum of the quantum-mechanical system~\eqref{Sch} can also be found from the Generalised Bohr-Sommerfeld quantisation condition (GBS),
\begin{equation}
    B(E) = \oint \limits_{\mathcal{C}_R} \qm(x, E) \s\d x = 2 \s \pi \s n\s\hbar\quad,\label{GBS}
\end{equation}
where
\begin{equation}\label{pivoisemki}
    \qm(x, E) = \dfrac{\hbar}{\iu} \dfrac{1}{\psi(x)} \dfrac{\d \psi(x)}{\d x}\quad,
\end{equation}
while the contour $\mathcal{C}_R$ encloses the classical turning points. The functions $\s\qm(x, E)\s$ and $\s B(E)\s$ are often referred to as the quantum momentum function (QMF) and the quantum action function (QAF), correspondingly. For an introduction to the generalised Bohr-Sommerfeld quantisation see Appendix~\ref{miauu}.

\subsection{Sextic QES potentials \label{sexticqes}}

In this section we discuss how the energy spectrum reflection (ESR) property of the sextic QES potential follows from the matrix form of the Hamiltonian. While~\cite{Shifman} addressed the case of the self-dual sextic potential, our analysis will naturally extend the discussion to two dual potentials, in analogy with the studies of the Lam\'e potential in~\cite{Dual}. Thereafter we explore how the ESR property can be formulated and proven in the language of the quantum action. Lastly, we discuss in greater detail the relation between the duality of classical and quantum actions.

Throughout this section we employ the units in which $\s\dfrac{\hbar^2}{2\s m} = \dfrac{1}{2}\s$.

\subsubsection{Algebraic approach \label{algappr}}

The most well-known examples of QES problems are the even and odd sixth-order polynomial potentials:~\footnote{~These potentials are discussed in~\cite{Shifman}. Be mindful, though, that the constant parts of these potentials are omitted there.}
\begin{subequations}\label{bothz}
\begin{gather}
    V_{\text{even}}(\nu,\mu,j,x) = \dfrac{\nu^2}{8}\s x^6 + \dfrac{\mu\s\nu}{4} \s x^4 + \left[  \dfrac{\mu^2}{8} - 2 \s \nu \left( j + \dfrac{3}{8} \right)\right]x^2 - \mu\left(j+\dfrac{1}{4}\right) \quad,\vspace{2mm}
    \label{even}\\
    V_{\text{odd}}(\nu,\mu,j,x) = \dfrac{\nu^2}{8}\s x^6 + \dfrac{\mu\s\nu}{4} \s x^4 + \left[  \dfrac{\mu^2}{8} - 2 \s \nu \left( j + \dfrac{5}{8} \right)\right]\s x^2  - \mu\left(j+\dfrac{3}{4}\right)\quad,\vspace{4mm}
    \label{odd}\\
    \nonumber
    \nu \in \mathbb{R}^+,\quad,\qquad \mu \in \mathbb{R} \quad,\qquad
    j = 1/2,\s 1,\s 3/2,\s 2,\s\ldots
\end{gather}\label{V}
\end{subequations}
\begin{sloppypar}
These potentials are even and odd in the following sense: for them, the first ${J = 2\s j + 1}$ lowest even (odd) energy levels can be obtained through diagonalisation of the tridiagonal matrices
\end{sloppypar}
\begin{equation}
\begin{gathered}
\mathcal{H}_{\text{``G'',\s even}}:\\
\begin{alignedat}{4}
    &\mathcal{H}_{k, k+1} & \ = \ & - k \s  (2\s k - 1) \\
    &\mathcal{H}_{k, k}   & \ = \ & \mu \s (k - j - 1)\\
    &\mathcal{H}_{k, k-1} & \ = \ & (k-2)\s \nu - 2 \s j \s \nu
\end{alignedat}
\end{gathered}
\qquad\qquad
\begin{gathered}
\mathcal{H}_{\text{``G'',\s odd}}:\\
\begin{alignedat}{4}
    &\mathcal{H}_{k, k+1} & \ = \ & - k \s  (2\s k + 1) \\
    &\mathcal{H}_{k, k}   & \ = \ & \mu \s (k - j - 1)\\
    &\mathcal{H}_{k, k-1} & \ = \ & (k-2)\s \nu - 2 \s j \s \nu
\end{alignedat}
\end{gathered}
\qquad
\begin{aligned}
k = 1,2,\ldots,J\quad.
\end{aligned}\label{HH}
\end{equation}
This is owing to the fact that, after an eigenvalue-preserving quasi-gauge transformation,$\s $\footnote{\begin{sloppypar*}~Quasi-gauge is a transformation of the form ${\s\psi(x) \to \psi_{\text{``G''}}(x)\s \me^{-a(x)}\s ,\s\s H\to \me^{-a(x)}\s H_{\text{``G''}}\s \me^{a(x)}\s }$. The function $a(x)$ is set real, wherefore the transformation is \textit{non-unitary}, and the norm of the wave function is not preserved. The physical meaning of such a transformation is straightforward: solutions to a one-dimensional Schr\"odinger equation are normally sought in the form of  ${\s\left(\s \sum_n a_n\s  x^n\s \right)\s \exp(.\s .\s .)\s}$ where the exponent defines the asymptotic behaviour. When the exponential is found, it remains to determine the prefactor ${\s \sum_n a_n\s  x^n\s}$. In special cases, this series truncates to a finite sum.\end{sloppypar*}
\par ~A key property of quasi-gauge transformations is that they preserve the eigenvalues of the operator. See~\cite{Shifman} for details.}
the Hamiltonians corresponding to potentials~\eqref{V} can be written as:
\begin{equation}
\begin{alignedat}{3}
    &H_{\text{``G'',\s even}} & \ = \ & -2\s T^{\s 0}\s T^{\s -} - (2j+1)\s T^{\s -} - \nu\s T^{\s +} + \mu\s T^{\s 0}\quad&,\\
    &H_{\text{``G'',\s odd}}  & \ = \ & -2\s T^{\s 0}\s T^{\s -} - (2j+1)\s T^{\s -} - \nu\s T^{\s +} + \mu\s T^{\s 0} -2\s T^{\s -}\quad&,
\end{alignedat}
\end{equation}
where the operators
\begin{equation}\begin{gathered}
    T^{\s +} = 2 \s j\s\xi - \xi^2 \s\dfrac{\d{}}{\d\xi} \quad,\qquad
    T^{\s 0} = - j + \xi \dfrac{\d{}}{\d\xi}\quad,\qquad
    T^{\s -} = \dfrac{\d{}}{\d\xi}\quad,\\ \xi=x^2\quad
\end{gathered}\end{equation}
obey the commutation relation:
\begin{equation}
    [T^{\s +},T^{\s -}] = 2\s T^{\s 0}\quad,\qquad [T^{\s +},T^{\s 0}] = -T^{\s +}\quad,\qquad[T^{\s -},T^{\s 0}] = T^{\s -}\quad,
\end{equation}
\begin{sloppypar}
and therefore can be interpreted as the generators of a representation of the $SU(2)$ group, with $j$ being the spin and ${J=2j+1}$ the representation dimension.
\end{sloppypar}

The potentials in~\eqref{bothz} can be obtained from the potential~\eqref{Vcl} by setting
\begin{subequations}
\begin{alignat}{22}
    a &= -\mu\left(j+\dfrac{1}{4}\right) \quad,\qquad
    b &&= \left[  \dfrac{\mu^2}{8} - 2 \s \nu\s  \left( j + \dfrac{3}{8} \right)\right] \quad,\qquad
    c &&= \dfrac{\mu\s \nu}{4} \quad,\qquad
    d &&= \dfrac{\nu^2}{8}
    \quad&&,\\
    \intertext{and}
    a &= -\mu\left(j+\dfrac{3}{4}\right) \quad,\qquad
    b &&= \left[  \dfrac{\mu^2}{8} - 2 \s \nu\s  \left( j + \dfrac{5}{8} \right)\right] \quad,\qquad
    c &&= \dfrac{\mu\s \nu}{4} \quad,\qquad
    d &&= \dfrac{\nu^2}{8}
    \quad&&,
\end{alignat}
\end{subequations}
correspondingly.
The change of the sign, $\mu \to - \mu$, leads to $\s \{a \to -a,\s b \to \s b,\s c \to -c,\s d \to d\}\s $, whence the potentials with opposite values of $\mu$ are dual.
In the notation
\begin{subequations}
\begin{alignat}{2}
    V_{\{ \mu \}} (x) &= V_{\text{even}}\s &(\nu,\mu,j,x) \quad,\\
    \widetilde{V}_{\{ \mu \}} (x) &= V_{\text{odd}}\s &(\nu,\mu,j,x) \quad.
\end{alignat}
\end{subequations}
the duality and self-duality conditions assume the form of
\begin{subequations}
\begin{alignat}{4}
    V_{\{ \mu \}}  (x)& = - V_{\{ - \mu \}}  (\iu \s x) \quad&,\qquad
    \widetilde{V}_{\{ \mu \}}  (x)& = - \widetilde{V}_{\{ - \mu \}} (\iu \s x) \quad&,
    \label{selfd1}
    \\
    V_{\{ 0 \}} \s (x)& = - V_{\{ 0 \}}  (\iu \s x)\quad&,\qquad
    \widetilde{V}_{\{ 0 \}}  (x)& = - \widetilde{V}_{\{ 0 \}}  (\iu \s x)
    \quad&.
    \label{selfd2}
\end{alignat}
\end{subequations}

In Appendix~\ref{prop}, we prove a simple lemma stating that the eigenvalues of a tridiagonal matrix change their signs when the signs of the diagonal elements are changed to their opposite. The two matrices~\eqref{HH} are exactly the case: they coincide, except for the diagonal elements. A diagonal element of the first matrix has a sign opposite to that of an analogous element of the second matrix.  So, in accordance with the said lemma, the energies from the algebraic parts of the spectra of the even and odd potentials change their signs under the switch $\mu \to -\mu$:
\begin{subequations}
\begin{alignat}{9}
    E_{\{\mu\},\s n} &= - E_{\{-\mu\},\s N-n} \quad&&,\qquad
    n=0,2,4\ldots,N
    \quad\quad\qquad \text{($\s $for~$\s V_{\text{even}}\s $)}
    \quad&&,\\
    \widetilde{E}_{\{\mu\},\s\widetilde{n}} &= - \widetilde{E}_{\{-\mu\},\s\widetilde{N} + 1 -\widetilde{n}} \quad&&,\qquad
    \widetilde{n}=1,3,5\ldots,\widetilde{N}
    \quad\quad\qquad \text{($\s $for~$\s V_{\text{odd}}\s $)}
    \quad&&.
\end{alignat}
\label{symmy}
\end{subequations}
Here $N$ and $\widetilde{N}$ denote the highest energy levels in the algebraic sectors of the potentials:
\begin{subequations}
\begin{alignat}{6}
    N &= 2\s J-2 &&= 4\s j \quad&&,\\
    \widetilde{N} &= 2\s J-1 &&= 4\s j + 1 \quad&&.
\end{alignat}
\end{subequations}

\subsubsection{Quantum action \label{QAsex}}

A different way to prove the ESR property~\eqref{symmy} of the energy levels of dual potentials follows from relating their quantum actions. This necessitates  deriving
a quantum analogue to~\eqref{S}. Despite the fact that we do not know the expression for the QMF (obtaining this would be equivalent to solving the Schr\"odinger equation), we can calculate its residue at infinity by substituting the Laurent expansion into the Ricatti equation, as was proposed in~\cite{LP}~---~see equation \eqref{ric2} in Appendix~\ref{miauu}.\s\footnote{~For the algebraic part of the spectrum of QES potentials, the solutions of the Schr\"odinger equation (and, thereby, the QMF) can, in principle, be obtained by means of the matrix algebra. Technically, though, this can be performed only if the size of the matrix, $J$, is small. Here we discuss a general method applicable to arbitrary potentials.}

As compared to the algebraic method, the current approach has an important advantage. While the algebraic method does not permit for studying the perturbative and WKB expansions, our approach offers this opportunity. Indeed, as we shall see shortly, much information on the perturbative and WKB expansions can be derived  from the exploration of the quantum action.

For the time being, we shall perform our calculations for the even potential. Thereafter, we shall provide the results for the odd potential, and shall briefly discuss the difference between these two situations.

We begin with noticing that from the Ricatti equation obeyed by the QMF,
\begin{equation}
    \qm^{\s 2}(x, E) + \dfrac{\hbar}{\iu}  \qm^{\s\prime}(x, E) = 2\s m\s(E - V(x)) \quad,\label{ric2text}
\end{equation}
and from the functional form of the potential~\eqref{even} it ensues  that the QMF has exactly the same symmetry property as the classical momenta:
\begin{equation}
    \qm_{\{-\mu\}}(x\s , E) = \iu\s \qm_{\{\mu\}}(\iu\s x,-E)\quad. \label{pdual}
\end{equation}

We will now explore how the generalised Bohr-Sommerfeld condition~\eqref{GBS} works over different ranges of energies. For energies below the maximum of the potential~$V_{\{\mu\}}(x)$,\s\footnote{~See Appendix~\ref{shapes} for a detailed explanation of all our assumptions.} this equation reads as
\begin{equation}\begin{gathered}
    \oint \limits_{\mathcal{C}_1} \qm_{\{\mu\}}(x, \s -E) \s \d x=
    \oint \limits_{\mathcal{C}_2} \qm_{\{\mu\}}(x, \s -E) \s \d x
    =2 \s  \pi\s \hbar \s m\quad,\\
    V_{\{\mu\},\s\text{min}} < -E < V_{\{\mu\}}(0) \quad.
    \label{GBSb}
\end{gathered}\end{equation}
As before, the contours are defined as in Figure~\ref{roots6}.

The symmetry property~\eqref{pdual} allows us to calculate the QAF in the equation~\eqref{GBS} for the positive energy levels of the potential $V_{\{-\mu\}}\s (x)$. The QAF is equal to an integral of $\qm_{\{\mu\}}\s (x,\s -E)$ along the imaginary axis:
\begin{equation}
    \oint \limits_{\mathcal{C}'_3} \qm_{\{-\mu\}}(x,\s  E) \d x =
    \oint \limits_{\mathcal{C}_3} \qm_{\{\mu\}}(x,\s  -E) \d x \quad.
\end{equation}
This observation becomes instrumental when we study the sum of the integrals of $\s \qm_{\{\mu\}} (x,\s -E)\s $ over the contours $\s \mathcal{C}_1\s $, $\s \mathcal{C}_2\s $ and $\s \mathcal{C}_3\s $, a sum which is equal to an integral over the large contour $\s \mathcal{C}_\infty\s $ enclosing \textit{all} the poles and branch cuts of $\s \qm_{\{\mu\}}(x,\s -E)\s $.
In \eqref{36} below, the integrals on the left-hand side are the QAFs for the potential $V_{\{-\mu\}}$ with the energy below the local maximum and for $V_{\{\mu\}}$ with the energy above the local maximum, respectively. The integral on the right-hand side can be evaluated with the aid of the residue of $\qm_{\{\mu\}} (x,\s -E)$ at infinity:
\begin{equation}
\begin{alignedat}{9}
    2 \oint \limits_{\mathcal{C}_1} \qm_{\{\mu\}} (x, \s -E) \d x +
    \oint \limits_{\mathcal{C}_3} \qm_{\{\mu\}} (x, \s -E)  \d x &=
    \oint \limits_{\mathcal{C}_\infty} \qm_{\{\mu\}} (x, \s -E)  \d x \\&=
    2\s \pi\s \iu\s \Res\{\qm_{\{\mu\}} (x, \s -E)dx, \infty\}\quad.
    \label{36}
\end{alignedat}\end{equation}
Here we have made use of the fact that the function $\s \qm_{\{\mu\}}(x,\s -E)\s $ has no poles other than those inside the contours $\s \mathcal{C}_1\s $, $\s \mathcal{C}_2\s $ or $\s \mathcal{C}_3\s $. Indeed, as was demonstrated in~\cite{LP}, the poles of the QMF emerge only on the branch cuts of the classical momentum, which in our case are the segments of the real and imaginary axes, enclosed by $\s \mathcal{C}_1\s $, $\s \mathcal{C}_2\s $ and $\s \mathcal{C}_3\s $.
This remarkable property allows us to employ the same integration contours for the QMF, as those employed for the classical momentum. We also assume the point $x=\infty$ to be an isolated singularity, which is indeed the case for the considered QES potential, as  was demonstrated in~\cite{HamJa}.

The equality (\ref{36}) establishes a relation between the quantisation condition for a system described by the potential $V_{\{\mu\}}(x)$, with the energy $-E$, and the quantisation condition for a system described by $V_{\{-\mu\}}(x)$, with the opposite energy $E$. In other words, starting from~\eqref{GBSb} we can derive the quantisation condition \eqref{GBS} for $V_{\{-\mu\}}(x)$.

The residue on the right-hand side of equation~\eqref{36} can be calculated directly from the Ricatti equation by expanding the QMF into the Laurent series, see Appendix~\ref{residuesh}. The result is:
\begin{equation}
    2 \oint \limits_{\mathcal{C}_1} \qm_{\{\mu\}}(x, \s -E) \d x +
    \oint \limits_{\mathcal{C}'_3} \qm_{\{-\mu\}}(x, \s E) \d x =
    \oint \limits_{\mathcal{C}_\infty} \qm_{\{\mu\}}(x, \s -E) \d x =
    8 \s  \pi \s j\s \hbar \quad
    \label{qduali}
\end{equation}
or, equivalently:
\begin{equation}
    2 \oint \limits_{\mathcal{C}_1} \qm_{\{\mu\}}(x, \s -E) \d x +
    \oint \limits_{\mathcal{C}'_3} \qm_{\{-\mu\}}(x, \s E) \d x =
    2\s\pi\s N\s\hbar
    \quad\quad\qquad \text{($\s $for~$\s V_{\text{even}}\s $)}
    \quad,\label{ALLeven}
\end{equation}
where
\begin{equation}
N = 2\s J-2 = 4\s j
\end{equation}
is the number of the highest energy level in the algebraic sector.
Moving the first integral in~\eqref{qduali} from the left to the right hand side, and then using the quantisation condition~\eqref{GBSb}, one arrives at
\begin{equation}
\begin{alignedat}{9}
    \oint \limits_{\mathcal{C}'_3} \qm_{\{-\mu\}}(x, \s E) \d x &=
    \oint \limits_{\mathcal{C}_\infty} \qm_{\{\mu\}}(x, \s -E) \d x
    - 2 \oint \limits_{\mathcal{C}_1} \qm_{\{\mu\}}(x, \s -E) \d x
    \\
    & =
    2\s \pi\s (N - 2\s m)\s \hbar
\end{alignedat}
    \quad\quad\qquad \text{($\s $for~$\s V_{\text{even}}\s $)}
    \quad,\label{GBSa}
\end{equation}
while~\eqref{ALLeven} can be written as
\begin{equation}
    2\s B_{\{\mu\}}(-E) + B_{\{-\mu\}}(E) = 2\s\pi\s N\s\hbar\quad.
    \label{ququdual}
\end{equation}

From looking at the formul\ae~\eqref{GBSb} and~\eqref{GBSa}, one may misconclude that there is a symmetry relation between the $m$-th energy level of the potential $V_{\{\mu\}}(x)$ and the $\s (N-2\s m)$-th energy level of $V_{\{-\mu\}}(x)$.
This, however, would be a mistake because there is a fundamental difference between the cases of two and four classical turning points for a given energy, i.e. motion above and below the local maximum.
In the former case, equation~\eqref{GBS} has a single solution for the energy, while in the latter case it usually possesses two solutions (an exception to this rule to be discussed shortly):
\begin{equation}
        \oint \limits_{\mathcal{C}_1} \qm_{\{\mu\}}(x, E) \d x = 2 \s  \pi \s m\s \hbar \qquad\rightarrow\quad
    \begin{cases}
        E_{\{\mu\},\s m} \qquad&\text{if} \qquad E_{\{\mu\},\s m\phantom{\s(+)}} > V_{\{\mu\}}(0) \\
        E_{\{\mu\},\s m(-)},\:E_{\{\mu\},\s m\s(+)} \qquad&\text{if}\qquad E_{\{\mu\},\s m\s (-)} < V_{\{\mu\}}(0)
    \end{cases}\quad.
    \label{solving}
\end{equation}

We intend to comply with a  standard convention wherein
(a) the notation $\s E_{\{\mu\},\,n}\s $ stands for the $n$-th energy level from the bottom, and (b) the bottom level corresponds to $\s n=0\s $. To that end, the energy levels below $V_{\{\mu\}(0)}$ should be enumerated as:
\begin{equation}
\begin{cases}
E_{\{\mu\},\s 2m} &= E_{\{\mu\},\s m\s (-)} \\
E_{\{\mu\},\s 2m+1} &= E_{\{\mu\},\s m\s (+)}
\end{cases}
\qquad \text{if}\quad
E_{\{\mu\},\s m\s (-)} < V_{\{\mu\}}(0)\quad.\label{numbering}
\end{equation}
\begin{sloppypar}
Strictly speaking,~\eqref{solving} does not cover all the possibilities. A situation may emerge, wherein the highest level below $V_{\{\mu\}}(0)$ is $E_{\{\mu\},\s m^*\s (-)}$, and it lacks a counterpart ${E_{\{\mu\},\s m^*\s (+)}}$. Nonetheless, in this case the proper numbering is ${E_{\{\mu\},\s 2m^*} = E_{\{\mu\},\s m^*\s (-)}}$, as is~\eqref{numbering}. See also the discussion in Appendix~\ref{exampleee}.
\end{sloppypar}

Now it becomes evident that, on the right-hand side of equation~\eqref{GBSa}, the quantity $2\s m$ is the number of the even energy level, $E_{\{\mu\},\s 2m}$.
In other words, for an even $n$, the $n$-th energy level is symmetric to the $(N-n)$-th energy level:
\begin{equation}
    E_{\{\mu\},\s n} = - E_{\{-\mu\},\s N-n} \quad,\qquad
    n=0,2,4\ldots,N
    \quad\quad\qquad \text{($\s $for~$\s V_{\text{even}}\s $)}
    \quad.\label{EReven}
\end{equation}

It is worth noting that the energies $\s E_{\{\mu\},\s m\s (-)}\s $ and $\s E_{\{\mu\},\s m\s (+)}\s $ share a perturbative part and differ only due to the non-perturbative splitting caused by tunneling effects. At the perturbative level (in neglect of splitting), one can say that the level $\s E_{\{-\mu\},\s N-n}\s $ is symmetric to two levels: $\s E_{\{\mu\},\s n}\s $ and $\s E_{\{\mu\},\s n+1}\s $.

For the odd potential~\eqref{odd}, integrating $\widetilde{\qm}_{\{\mu\}}(x, \s -E)$ over the contour $\mathcal{C}_\infty$ gives:
\begin{equation}
    \oint \limits_{\mathcal{C}_\infty} \widetilde{\qm}_{\{\mu\}}(x, \s -E) \d x
    = 2\s \pi\s (\widetilde{N}+1)\s \hbar
    \quad\quad\qquad \text{($\s $for~$\s V_{\text{odd}}\s $)}
    \quad,
    \label{ALLodd}
\end{equation}
where the number of the highest energy level in the algebraic sector is now given by:
\begin{equation}
\widetilde{N} = 2\s J-1 = 4\s j + 1 \quad.
\end{equation}
Hence, for the quantum action we may write:
\begin{equation}
    \widetilde{B}_{\{\mu\}}(E) + 2\s \widetilde{B}_{\{-\mu\}}(-E) = \ 2\s \pi\s (\widetilde{N}+1)\s \hbar
    \quad.
    \label{quququdual}
\end{equation}
Then, starting again with~\eqref{GBSb}, we arrive at:
\begin{equation}
    \oint \limits_{\mathcal{C}'_3} \widetilde{\qm}_{\{-\mu\}}(x, \s E) \d x =
    2\s \pi\s (\widetilde{N}  - 2\s m)\s \hbar =
    2\s \pi\s \left(\widetilde{N}  + 1 - (2\s m + 1)\right)\hbar
    \quad\quad\qquad \text{($\s $for~$\s V_{\text{odd}}\s $)}
    \quad.\label{GBSaodd}
\end{equation}
Be mindful that in the case of the odd potential, one has to choose $\widetilde{E}_{\{\mu\},\s m\s (+)} = \widetilde{E}_{\{\mu\},\s 2m+1}$ when solving~\eqref{solving}. Therefrom the energy spectrum reflection reads:
\begin{equation}
    \widetilde{E}_{\{\mu\},\s\widetilde{n}} = - \widetilde{E}_{\{-\mu\},\s\widetilde{N} + 1 -\widetilde{n}} \quad,\qquad
    \widetilde{n}=1,3,5\ldots,\widetilde{N}
    \quad\quad\qquad \text{($\s $for~$\s V_{\text{odd}}\s $)}
    \quad.\label{ERodd}
\end{equation}
Importantly, the ESR symmetry not only implies the duality of the numerical values of the corresponding energy levels, but it also implies the duality of the WKB expansions generated by the generalised Bohr-Sommerfeld quantisation condition.
To appreciate this, let us take a look at equations~\eqref{GBSb} and~\eqref{GBSa}. While the former generates the WKB expansion for the low levels, the latter generates the WKB expansion for the dual levels. Equation~\eqref{ALLeven} shows that these expansions should, up to a constant, match term by term. If we also recall that the generalised Bohr-Sommerfeld quantisation condition can be used for generating a perturbative expansion we immediately recognise that the perturbative expansion of the low levels has to match with the WKB expansion of the dual excited levels, an observation first made in~\cite{Shifman}.

\subsubsection{WKB approach \label{wkbappr}}

Now we want to trace the connection between the duality property of the classical action~\eqref{S} and its quantum analogues~\eqref{ququdual} and~\eqref{quququdual}. To do so we employ the WKB method and show that equations~\eqref{EReven} and~\eqref{ERodd} can be obtained in the Bohr-Sommerfeld approximation (i.e., by taking into account the leading-order and next-to-leading-order terms in the expansion of the QMF in powers of $\s \hbar\s $). This however is not a general rule. In the next section we perform an analogous calculation for a periodic potential. As we will see, in that case it is necessary to take into account \textit{all} perturbative orders of the WKB expansion, in order to reconstruct the exact ESR property.

We begin by writing down the classical duality condition~\eqref{S} in the notation introduced in~\eqref{V}. For the even potential we arrive at:
\begin{equation}\begin{gathered}
    \oint \limits_{\mathcal{C}_\infty} p_{\{\mu\}}(x,\s -E)\d x =
    2\s \pi\left(4\s j + \dfrac{3}{2}\right)\hbar =
    2\s \pi\left(N + \dfrac{3}{2}\right)\hbar
    \quad,\label{symBS}
\end{gathered}\end{equation}
which coincides with~\eqref{ALLeven} except for the extra term $\s \dfrac{3}{2}\s $ in the brackets. This term is a sum of the Maslov indices of the two potentials which is derived from the order-$\hbar$ correction to the classical action. Indeed, the insertion of the series expansion~\eqref{L8} into the RHS of~\eqref{36} yields:~\footnote{~In Appendix~\ref{miauu} we mention that, beginning from $k=2$, it is only the even terms that appear on the LHS of~\eqref{sumcancel1}.}
\begin{equation}
\begin{gathered}
    2 B_{\{\mu\}}(-E) + B_{\{-\mu\}}(E) =
    2 \s \pi\s \iu\s \Res \{ \qm_{\{\mu\}}(x,\s -E)dx,\s\infty \} =
    \oint \limits_{\mathcal{C}_\infty} \sum \limits_{k=0}^{\infty} \left(\dfrac{\hbar}{\iu}\right)^k \qm_{\{\mu\}\s k}(x,\s -E) \d x \\=
    \oint \limits_{\mathcal{C}_\infty} p_{\{\mu\}}(x,\s -E)\d x
    + \dfrac{\hbar}{\iu} \oint \limits_{\mathcal{C}_\infty} \qm_{\{\mu\}\s 1}(x,\s -E) \d x
    + \left(\dfrac{\hbar}{\iu}\right)^2 \oint \limits_{\mathcal{C}_\infty} \qm_{\{\mu\}\s 2}(x,\s -E) \d x +
    \ldots
    \\= 2\s \pi\s \iu\s \Bigg(
    \underbrace{ \vphantom{\left(\dfrac{\hbar}{\iu}\right)^2}
    \Res\left\{ p_{\{\mu\}}(x,\s -E)dx,\s\infty \right\}
    }_{\text{The classical result, eq.~\eqref{symBS}}} +
    \underbrace{ \dfrac{\hbar}{\iu}\s  \vphantom{\left(\dfrac{\hbar}{\iu}\right)^2}
    \Res\left\{ \qm_{\{\mu\}\s 1}(x,\s -E)dx,\s\infty \right\}
    }_{\text{The sum of three Maslov indices}}
    + \\
    \underbrace{ \left(\dfrac{\hbar}{\iu}\right)^2
    \Res\left\{ \qm_{\{\mu\}\s 2}(x,\s -E)dx,\s\infty \right\} + \ldots
    }_{\text{Vanishes for the sextic potential}}
    \Bigg)
    = 2\s \pi\s \left(N + \dfrac{3}{2}\right)\hbar - 3\s \pi\s \hbar
    = 2\s\pi\s n\s\hbar \quad.
\end{gathered}
\label{sumcancel1}
\end{equation}
Thus, due to the unique feature of the considered QES potential,
\begin{equation}
     \Res\{ \qm_{\{\mu\}\s k}(x,\s -E)dx, \infty \} = 0 \qquad\text{for}\qquad k \geq 2 \quad,\label{res0}
\end{equation}
the quantum ESR symmetry can be explained in the Bohr-Sommerfeld approximation. The quantum corrections to the momentum function have no residue at $x=\infty$, hence their effects on the action functions are equal but opposite above and below the potential maximum.

Had we carried out our proof starting from the usual Bohr-Sommerfeld quantisation condition $S(E)=2\pi\hbar(n+1/2)$ instead of~\eqref{GBS}, we would be able to prove the ESR symmetry at the perturbative level only. In other words, we would have proved that in the double-well energy region each pair of levels (those equal in neglect of tunneling) is symmetric to an energy level in the single-well region of another potential.

\subsubsection{Non-perturbative effects \label{nonpert}}

We now offer a couple of comments on  non-perturbative corrections which never showed up in our developments.
As one calculates the energy in potentials with degenerate minima, such corrections typically arise:
(a) below the absolute maximum of the potentials, due to the tunneling between the two wells; (b) above the absolute maximum (in the single-well region), owing to the over-the-barrier reflection. In the case of deep wells, these corrections become exponentially small for energy levels far from the maximum of the potential.



It would be important to undescore the following two circumstances:

\begin{enumerate}
\item
Up to equation
\eqref{sumcancel1}, no non-perturbative corrections have been omitted, since we had made no assumption on the specific form of the QMF. For this reason, the equations~\eqref{EReven} and~\eqref{ERodd} are \textit{exact}~---~a fact already known from the algebraic considerations, see Section~\ref{algappr}).

\item As we already stated, the form of the series in~\eqref{sumcancel1} implies neglecting the non-perturbative terms. The more remarkable it is that we have nevertheless arrived at the exact result~\eqref{sumcancel1}.
A possible explanation is as follows.
The instanton action (calculated over a cycle enclosing $BC$ in Figure~\ref{roots6}) and the action emerging due to the over-the-barrier reflection ($B^{\s \prime\s }C^{\s \prime}$ in Figure~\ref{roots62}) have opposite signs and cancel in~\eqref{sumcancel1}.
Accordingly, the corrections to the symmetric energy levels have opposite signs. The exponentially small leading-order corrections are easy to obtain, and they indeed do match. The duality property of the potential ensues that
\begin{equation}\label{eq:tunneling}\begin{alignedat}{9}
    &\Delta_{-\epsilon} &&\propto - &&\exp\left(- \dfrac{2}{\hbar} \int \limits_B^C p_{\{\mu\}}(x,\s -E)dx \right) \quad&&,\\
    &\Delta_{\epsilon} &&\propto &&\exp\left(- \dfrac{2}{\hbar} \int \limits_{B'}^{C'} p_{\{-\mu\}}(\iu \s x,\s E)dx \right)  \sim -\Delta_{-\epsilon} \quad&&.
\end{alignedat}\end{equation}
\end{enumerate}

\subsection{Lam\'e potential \label{ellqes}}


In this section we discuss the Lam\'e potential, a periodic QES potential, and perform calculations analogous to those done in Section~\ref{sexticqes} for the sextic potentials. First, we briefly summarise the results of the algebraic approach elaborated in~\cite{Dual}.
Then, we perform an analysis similar to that in Section~\ref{QAsex}, i.e., we find an exact expression that interrelates the QAFs of dual potentials and explore their WKB expansion.

Throughout section~\ref{ellqes} we use units in which $\s\dfrac{\hbar^2}{2\s m} = 1\s$.

\subsubsection[Algebraic approach]{Algebraic approach
}

The widely studied~\cite{ars,akhi,susy,Dual} potential of the Lam\'e model has the form
\begin{equation}
\Vl(x|\nu) = J(J+1)
\left( \nu \snt(x|\nu) - \dfrac{1}{2} \right)\quad.
\label{LAME}
\end{equation}
It can be obtained from the potential~\eqref{Ve} by setting
\begin{equation}\label{knock}
a = J\s(J+1) \quad,\quad b = -\dfrac{1}{2} J(J+1) \quad,\qquad J =1,\s 2,\s 3,\s\ldots
\end{equation}
For uniformity of terminology we also refer to this potential as \textit{elliptic} (as we did in the classical case). The corresponding Schr\"odinger equation,
\begin{equation}
    \label{schlam}
    \left[-\dfrac{\d{}^2}{\d x^2} + \Vl(x|\nu)\right] \psi(x) = E\s \psi(x)\quad,
\end{equation}
\begin{sloppypar*}
has a continuous spectrum comprising $J$ finite bands and an upper band extending to infinity. For the potential $V(x|\nu)$ we label the band edges as $E_l [\nu]$, with ${l=1,\s 2,\s\ldots,\s (2\s J+1)}$. In this notation, the band-gap structure of the potential reads
\end{sloppypar*}
\begin{equation}\begin{alignedat}{12}
        &\text{Bands:}\qquad &E_{2k-1} &\leq &E &\leq E_{2k}  \quad&\text{and}\quad &E\geq E_{2k+1}\\
        &\text{Gaps:} \qquad &E_{2k}   &\leq &E &\leq E_{2k+1}\quad&\text{and}\quad &E\leq E_{1}
\end{alignedat}
\quad,\qquad k=1,\s 2\s \ldots J \quad.
\end{equation}
The Lam\'e potential is exactly solvable in the sense that its band-edge energies can be found as eigenvalues of the operator~\cite{Dual}
\begin{equation}
H[\nu] = J^2_x + \nu\s J^2_y - \dfrac{1}{2}\s J\s (J+1)\s \unit \quad.
\label{lamop}
\end{equation}
Here $\s \unit\s $ is a unit matrix,
while $\s J_k\s $ are  generators of the $\s SU(2)\s $ group in the spin $J$ representation, obeying the usual commutation relation $\s {[J_x,\s J_y] = \iu\s J_z}\s $.

The ESR property
\begin{equation}
    E_l[\nu] = - E_{2J+2-l} [1-\nu]
    \label{ERlamee}
\end{equation}
immediately ensues from
\begin{equation}
H[\nu] = J^2_x + \nu\s J^2_y - \dfrac{1}{2}\s J\s (J+1)\s \unit =
- \left[ \left(J^2_z + (1-\nu)\s J^2_y\right) - \dfrac{1}{2}\s J\s (J+1)\s \unit\right]
\quad,
\label{lamopdual}
\end{equation}
because the operator $\s \left[ J^2_z +(1-\nu)\s J^2_y \right]\s $ has the same eigenvalues as $\s \left[ J^2_x +(1-\nu)\s J^2_y \right]\s $.\s\footnote{~
The RHS of equation \eqref{lamopdual} can be rewritten as $\s-R H[1-\nu]R^\dagger\s$, where $R$ is a Hermitian matrix. Clearly, such an operator has the same eigenvalues as $\s -H[1-\nu]\s$.
}

In the next section, we study the duality property in the language of the quantum action.

\subsubsection{Quantum action}

We will now show that the QAF of the Lam\'e equation possesses a duality property similar to~\eqref{ququdual} and~\eqref{quququdual} for the sextic potential. An equation defining the QMF for the Lam\'e potential comes up through the insertion the potential~\eqref{LAME} into the Ricatti equation~\eqref{ric2text}. The symmetry property of the QMF has the same shape as \eqref{plame} for the classical momentum:
\begin{equation}\begin{aligned}
\qm(\iu\s x+K+\iu\s K',\s\Vl_{\nu,\s\text{max}}-\epsilon\s |\s \nu) =
\iu\s \qm(x,\s\Vl_{1-\nu,\s\text{max}}+\epsilon\s |\s 1-\nu) \quad.
\end{aligned}
\end{equation}
Now, taking into account that
\begin{equation}
    \Vl_{\nu,\s\text{max}} = \dfrac{1}{2}\s J\s (J+1)\s (-1+2\s \nu) = - \Vl_{1-\nu,\s\text{max}} \quad,
\end{equation}
and changing
\begin{equation}
    \epsilon \to \epsilon + \Vl_{\nu,\s\text{max}} \quad,
\end{equation}
we arrive at
\begin{equation}\begin{aligned}
\qm(\iu\s x+K+\iu\s K',\s -\epsilon\s |\s \nu) =
\iu\s \qm(x,\s\epsilon\s |\s 1-\nu) \quad.
\end{aligned}
\label{qplame}
\end{equation}
The latter equality looks very much like equation~\eqref{pdual} for the sextic potential, and leads us to similar results.

In terms of the quantum actions, the ESR property~\eqref{ERlamee} becomes
\begin{subequations}\label{dudu}
\begin{alignat}{5}
    B(-\epsilon\s |\s \nu) &= \oint\limits_{\mathcal{C}_c} \qm(x,\s -\epsilon\s |\s \nu)dx &&= 2\s \pi(J-n)\s \hbar \quad&&,\label{dudu1}\\
    B(\epsilon\s |\s1- \nu) &= \oint\limits_{-\mathcal{C}_2} \qm(x,\s\epsilon\s |\s 1-\nu)dx &&= \pi\s n\s \hbar \quad&&.\label{dudu2}
\end{alignat}
\end{subequations}
It then can be shown that the {$n$-th} band edge of the potential ${V^{\text{L}}(x|1-\nu)}$, as determined from equation~\eqref{dudu2}, is the negative of the {$(J-n)$-th} band edge of the potential ${V^{\text{L}}(x|\nu)}$, which is found from equation~\eqref{dudu1}.
To prove this, we demonstrate that the formul\ae~\eqref{dudu1} and~\eqref{dudu2} are equivalent and may be derived from one another.
For example, let us assume the validity of~\eqref{dudu2} and demonstrate how it entails~\eqref{dudu1}.

The integration contours for the QMF are the same as in the classical case, see Figure~\ref{pxnuriem}.
The LHS of~\eqref{dudu1} can be expressed as:
\begin{equation}\begin{alignedat}{2}
    \oint \limits_{\mathcal{C}_c} \qm(x,\s -\epsilon\s |\s \nu)dx &=
    \oint \limits_{-\mathcal{C}_p} \qm(x,\s -\epsilon\s |\s \nu)dx
     +2\s  \oint \limits_{\mathcal{C}_2} \qm(x,\s\epsilon\s |\s 1-\nu)dx \\
    &= - 2\s \pi\s \iu \s  \Res\{\qm(x,\s -\epsilon\s |\s \nu)dx,\s\iu\s K'\}
    - 2\s  \oint \limits_{-\mathcal{C}_2} \qm(x,\s\epsilon\s |\s 1-\nu)dx \quad.
    \label{snres}
\end{alignedat}\end{equation}
The residue of the QMF at its pole is calculated by expanding it into the Laurent series and plugging the result into the Ricatti equation:
\begin{alignat}{2}
    - 2\s \pi\s \iu\s \Res\{\qm(x,\s -\epsilon\s |\s \nu)dx,\s\iu\s K'\} =
    2\s \pi\s J\s \hbar\quad.
    \label{sokolov}
\end{alignat}
The insertion of the equations \eqref{dudu2} and (\ref{sokolov}) in equation (\ref{snres}) yields:
\begin{equation}
        \oint \limits_{\mathcal{C}_c} \qm(x,\s -\epsilon\s |\s \nu)dx =
        2\s \pi\s J\s \hbar - 2\s (\pi\s n\s \hbar) =
        2\s \pi\s (J - n)\s \hbar\quad,
        \label{snansw}
\end{equation}
while equation~\eqref{snres} can be written as:
\begin{equation}\label{snfullres}
    B(-\epsilon\s |\s \nu) + 2\s B(\epsilon\s |\s 1-\nu) = 2\s \pi\s J\s \hbar\quad.
\end{equation}
Thus, in full analogy with the case of the sextic potential, equations~\eqref{dudu} and~\eqref{snfullres} explain the matching of the perturbative and WKB expansions of dual band edges~\cite{Dual}.

To understand the physical meaning of the index $\s n\s $, recall that equation~\eqref{dudu} provides the band-edges $\s E_{2n+1}\s $ and $\s E_{2n+2}\s $. In situations where non-perturbative terms can be neglected, this equation fully defines the \textit{locations of bands and gaps}.

\subsubsection{WKB approach}

Now we study, order by order in~$\hbar$, the duality property of the QAF, expressed by equation~\eqref{snfullres}. At the level of the ordinary Bohr-Sommerfeld quantisation one gets:
\begin{equation}\begin{alignedat}{9}
    S(-\epsilon\s |\s \nu) + 2\s S(\epsilon\s |\s 1-\nu)&=
    -2\s \pi\s \iu\s \Res \{ p^{\text{L}}(x,-\epsilon\s |\s \nu)dx,\s\iu\s K' \} \\
    = \oint \limits_{-\mathcal{C}_p} p^{\text{L}}(x,\s -\epsilon\s |\s \nu) \d x &= 2\s \pi\s \sqrt{J\s (J+1)}\s \hbar
    =2\s \pi\left( J + \dfrac{1}{2} -\dfrac{1}{8\s J} + \ldots \right)\hbar
    \quad,
\end{alignedat}\end{equation}
which in the limit $J\to \infty$ differs from the exact equality \eqref{snfullres} by the Maslov index $\dfrac{1}{2}$. For convenience, we introduce the notation.
\begin{equation}
    \kappa = \sqrt{J\s(J+1)} \quad.
\end{equation}
Taking into consideration the higher-order WKB terms in~\eqref{snfullres} yields:
\begin{equation}
\begin{gathered}
    B(-\epsilon\s |\s \nu) + 2 B(\epsilon\s |\s 1-\nu) =
        -2\s \pi\s \iu\s \Res \{ \qm(x,-\epsilon\s |\s \nu),\s\iu\s K' \} =
    \oint \limits_{-\mathcal{C}_p} \sum \limits_{k=0}^{\infty} \left(\dfrac{\hbar}{\iu}\right)^k \qm_k(x,\s -\epsilon\s |\s \nu) \d x \\
    \\
    =
    \oint \limits_{-\mathcal{C}_p} p^{\text{L}}(x,\s -\epsilon)\d x
    + \dfrac{\hbar}{\iu} \oint \limits_{-\mathcal{C}_p} \qm_{1}(x,\s -\epsilon\s |\s \nu)\d x
    + \left(\dfrac{\hbar}{\iu}\right)^2 \oint \limits_{-\mathcal{C}_p} \qm_{2}(x,\s -\epsilon\s |\s \nu)\d x +
    \ldots
    \\
    \\
    =
    -2\s \pi\s \iu
    \left[
    (\iu\s  \kappa\s \hbar) +
    \dfrac{\hbar}{\iu} \left(\dfrac{1}{2}\right) +
    \left(\dfrac{\hbar}{\iu}\right)^2 \left( \dfrac{1}{8\s \iu\s \hbar\s \kappa} \right) +
    \left(\dfrac{\hbar}{\iu}\right)^3 ( 0 ) +
    \left(\dfrac{\hbar}{\iu}\right)^4 \left( \dfrac{1}{128\s \iu\s \hbar^3\s \kappa^3} \right) + \ldots
    \right]
    \\
    \\
    \hspace{5em}
    =
    2\s \pi\s \hbar
    \left[
    \kappa - \dfrac{1}{2}  + \dfrac{1}{8\s \kappa} - \dfrac{1}{128\s \kappa^3} + \ldots
    \right]
    = 2\s \pi\s \hbar\s  \dfrac{\sqrt{4\s \kappa^2+1}-1}{2}
    = 2\s \pi\s J\s \hbar
    \quad.
\end{gathered}
\label{sumcancel2}
\end{equation}
We see that summation of all (perturbative) orders of the QAF reproduces the exact equality~\eqref{snfullres}.
The explicit form of the higer-order terms in the WKB expansion of the QMF can be used to perform the calculation of the residues, see equation~\eqref{qmkp}. We remind that all the odd terms starting from $k=3$ vanish
(and the corresponding residues have to be equal to zero).

\begin{sloppypar}
In other words, within the Bohr-Sommerfeld approximation, by having started from~\eqref{dudu2} we would end up with ${\s2\s \pi\s ( \sqrt{J\s (J+1)} - 1/2 - n)\s \hbar\s}$ on the RHS of~\eqref{dudu1}.
\end{sloppypar}

The above calculation gives the correct result, even though this calculation does not account for non-perturbative effects. The reason for this is the same as was in the case of the double-well potential: the integration periods corresponding to tunneling and over-the-barrier reflection have opposite signs and cancel each other in equation~\eqref{sumcancel2}, making the so-obtained result exact.

\section{Summary and discussion}
The notion of duality, in the sense in which it is used in the current paper, was first introduced in~\cite{Dual,Shifman,ShifmanTurbiner} as a symmetry of the spectra of quantum-mechanical systems. Largely inspired by~\cite{HamJa}, we came up with an idea of describing such systems in terms of the quantum action function. We have demonstrated that for the sextic and  Lam\'e potentials this action possesses a symmetry analogous to that of the classical action. For this reason we introduced the notion of duality of the action in classical systems, see equations~\eqref{S} and~\eqref{Ss}. Generally, we expect equalities of this kind to hold for the classical counterparts of all quantum systems having ESR symmetry of quantum energy levels.


We have demonstrated that the duality property of the quantum action function of the sextic quasi-exactly solvable potential and the Lam\'e potential reads as:
\begin{subequations}
\begin{alignat}{10}
    &B_{\{\mu\}}(E) &&+ 2B_{\{-\mu\}}(-E) &&= 2B_{\{\mu\}}(-E) &&+ B_{\{-\mu\}}(E) & \ = \ & 2\s\pi\s N\s\hbar
    \quad\quad &&\text{($\s $for~$\s V_{\text{even}}\s $)}    \quad&,\label{resu1}\\
    &\widetilde{B}_{\{\mu\}}(E)&& + 2\widetilde{B}_{\{-\mu\}}(-E) &&= 2\widetilde{B}_{\{\mu\}}(-E) &&+ \widetilde{B}_{\{-\mu\}}(E) & \ = \ & 2\s \pi\s (\widetilde{N}+1)\s \hbar
    \quad\quad &&\text{($\s $for~$\s V_{\text{odd}}\s $)}     \quad&,\label{resu2}\\
    &B(-\epsilon\s |\s \nu) &&+ 2 B(\epsilon\s |\s 1-\nu) &&= 2 B(-\epsilon\s |\s \nu) &&+ B(\epsilon\s |\s 1-\nu) & \ = \ & 2\s \pi\s J\s \hbar
    \quad\quad &&\text{($\s $for~$\s V^L\s $)}     \quad&,\label{resu3}
\end{alignat}
\label{resu}
\end{subequations}
where $B$ is the quantum action function. $J$ stands for the number of bands for the elliptic potential, while $N$ and $\widetilde{N}$ denote the numbers of the highest energy levels in the algebraic sector for the even and odd sextic potentials, correspondingly.
We have pointed out that the duality condition, when written in the above form, explains the matching between perturbative and WKB expansions of the dual energy levels, which was observed earlier in~\cite{Dual} and~\cite{Shifman}, see our equations~\eqref{ALLeven} and~\eqref{GBSa}, \eqref{ALLodd} and~\eqref{GBSaodd}, \eqref{dudu1} and \eqref{dudu2}.

For a particular form of the sextic quasi-exactly solvable potential, we discussed in great detail how the number of the energy levels is related to the integer on the right-hand side of the generalised Bohr-Sommerfeld quantisation condition, see Appendix~\ref{exampleee}. We furthermore extended the notion of self-duality to duality, in analogy to the case of the Lam\'e potential.

In our calculations, the non-perturbative effects  never appear explicitly. They are hidden inside the quantum action function which contains all the information about the quantum-mechanical system. Therefore, within the non-perturbative treatment~--- i.e., in the language of $\qm(x,E)\,$~--- the non-perturbative effects are automatically taken into account. If one first proves the duality in a perturbative way~--- i.e., in terms of $\qm_k(x,E)$~--- then one will see that the non-perturbative effects shift the dual energy levels in  opposite directions and therefore preserve the ESR property. For the first-order tunneling this can be easily seen from the symmetry of the classical momentum, see equation \eqref{eq:tunneling}.

The two key properties of the considered potentials, from which the duality of the classical action can be derived, are: (a)  energy-independence of the residues of the momentum at its poles; (b) the absence of additional branch points in the complex plane, except for the turning points along the real and imaginary axes.\s\footnote{~In principle, having additional poles with \textit{energy-independent} residues would also be compatible with duality.} We believe that these conditions are sufficient for a potential to possess the ESR symmetry.

Investigation of the algebraic and topological properties of the  Riemann surface of the classical momentum allowed us to derive the ESR symmetry at the Bohr-Sommerfeld level. We also explained briefly how the classical action can be found by means of the Picard-Fuchs equation (more detailed calculations to be presented in the upcoming paper~\cite{SHORT}\,). Further, we considered the WKB expansions of the formul\ae~\eqref{resu} and explored how these equalities  hold in each consecutive order of $\s \hbar\s$. From this consideration, we deduced that the ESR symmetry holds at any order in the WKB series for both potentials. The key to this finding is that
\textit{all} terms $\qm_k(E)$ in the perturbative expansion of the quantum momentum function are 1-forms defined on the same Riemann surface,
 and that all these forms have vanishing residues at the poles. In the upcoming paper \cite{SHORT} we will also present a straightforward extension of the Picard-Fuchs approach to calculate the quantum corrections.


\section*{{Acknowledgements}\label{thanks}}
\addcontentsline{toc}{section}{\nameref{thanks}}
The authors are grateful to Misha Shifman for suggesting the problem, and to Peter Koroteev whose consultation and advice were crucial for the progress on this project. M.K. would like to thank Vladimir Bychkov and Yevhen Kurianovych for fruitful discussions, and Michael Efroimsky for his continuous support.

\begin{appendices}

\section{Generalised Bohr-Sommerfeld quantisation condition\label{miauu}}

Here we provide a brief overview of facts from Quantum Mechanics, which we use in our study. Our starting point is the Schr\"odinger equation in one dimension:

\begin{equation}
    \widehat{H} \s \psi(x) = E \s \psi (x)
    \quad,\qquad
    \widehat{H} = \dfrac{\widehat{p}^{\s 2}}{2\s m} + V(x)
    \quad.
\end{equation}
Performing the substitution
\begin{equation}
    \psi(x) = \exp\left( \iu\s\sigma(x, E) / \hbar \right) \quad,\label{ch}
\end{equation}
{we observe that the function $\sigma^{\s\prime}(x,E) \equiv \dfrac{\d\sigma(x,E)}{\d x}$ satisfies the Ricatti equation:}
\begin{equation}
    (\sigma^{\s\prime}(x, E))^2 + \dfrac{\hbar}{\iu} \sigma^{\s\prime\prime}(x, E) = 2\s m\s(E - V(x)) \quad.\label{ric1}
\end{equation}
It ensues from this equation that in the limit of $\hbar \to 0$ the function \begin{equation}
    \qm(x, E) \equiv  \sigma^{\s\prime}(x, E)\quad.
    \label{290}
\end{equation}
satisfies the equation for the classical momentum. So $\qm(x, E)$ is termed the \textit{quantum momentum function} (QMF).

Accordingly, the Ricatti equation~\eqref{ric1} takes the form:
\begin{equation}
    \qm^{\s 2}(x, E) + \dfrac{\hbar}{\iu}  \qm^{\s\prime}(x, E) = 2\s m\s(E - V(x)) \quad.\label{ric2}
\end{equation}

The quantisation condition, whence the $n$-th energy level is determined, is normally obtained from the requirement of single-valuedness of the function $\psi(x)$. However, in~\cite{HamJa} a more interesting option was proposed. It was based on the fact that the wave function corresponding to the $n$-th energy level has $n$ zeros on the real axis, between the classical turning points (the latter points being the zeros of the classical momentum)~\cite{LL}.

In these zeroes, the QMF  has poles. Indeed, it trivially follows from (\ref{ch}) and (\ref{290}) that
\begin{equation}
    \qm(x, E) = \dfrac{\hbar}{\iu} \dfrac{1}{\psi(x)} \dfrac{\d \psi(x)}{\d x}\quad.
\end{equation}
For analytic potentials the pole of the function $\qm(x,E)$ is of first order, and the residue at this pole is $(-\iu \hbar)$. Therefore the integral of the QMF along the contour $\mathcal{C}_R$ enclosing classical turning points is
\begin{equation}
    B(E) = \oint \limits_{\mathcal{C}_R} \qm(x, E) \s\d x = 2 \s \pi \s n\s\hbar\quad,\label{GBSAppendix}
\end{equation}
where the contour $\mathcal{C}_R$ should be close enough to the real axis, in order to avoid containing the poles and branch cuts of $\qm(x,E)$, that are off the real axis. In the classical limit ($\hbar \to 0$), the series of poles inside $\mathcal{C}_R$ coalesces into a branch cut of the classical momentum~\cite{LP}.

The function $B(E)$ is sometimes referred to as the quantum action function (QAF)~\cite{LP,LP2}, and the equality~\eqref{GBSAppendix} itself as the Generalised Bohr-Sommereld quantisation condition (GBS). The GBS is often employed as a starting point in studies of the spectra of quantum systems.
The two common approaches to extract information from the Ricatti equation are:
\begin{enumerate}
\item To employ the expansion of the QMF in powers of $\hbar$ (the WKB method):
\begin{equation}
    \qm(x, E) = \sum \limits_{k=0}^{\infty} \left(\dfrac{\hbar}{\iu}\right)^k \qm_k(x, E) \quad.
    \label{L8}
\end{equation}
A strong side of this approach lies in its clear physical meaning, in that the first two terms give the ordinary Bohr-Sommerfeld quantisation condition. A disadvantage of this method is that the form of~\eqref{L8} implies the neglect of exponentially small non-perturbative terms. Hence all the results are valid only when tunneling is unimportant (i.e., at a distance from the relative maxima of the potential). This drawback can, however, be overcome, if one begins looking for the quantum action (and, thereby, the energy) in form of a \textit{trans-series} in $\hbar$ rather than a polynomial. Moreover, until recently, the GBS used to be the only tool to find the multi-instanton corrections to the energy in quantum mechanics~\cite{Mult1}.\footnote{~Several years ago, Dunne and \"{U}nsal~\cite{UWKB} suggested to build resurgent expansions by the so-called Uniform WKB Method. For some QM potentials, that method provides a more efficient way of calculating these exponentially small corrections to the energy. It also reveals some other interesting properties of such potentials.}

Substituting~\eqref{L8} into the Ricatti equation~\eqref{ric2} gives a recursive relation
\begin{gather}
    \sum \limits_{l=0}^{k} \qm_l(x,E)\s \qm_{k-l}(x,E) + \qm_{k-1}'(x,E) = 0 \quad,
\end{gather}
which allows to express all higher terms~\eqref{L8} through the classical momentum:
\begin{gather}\begin{gathered}
    \qm_k(x,E) = - \dfrac{1}{2\s \qm_0(x,E)} \left(\qm'_{k-1}(x,E)+\sum \limits_{l=1}^{k-1} \qm_l(x,E) \s \qm_{k-l}(x,E) \right)\quad,\\
    \qm_0 (x,E) = p(x,E)\quad.
    \label{qmkp}
\end{gathered}\end{gather}

We define the $k$-th correction to the classical action as
\begin{equation}
 \sigma_k (E) = \left(\dfrac{\hbar}{\iu}\right)^k \int \limits_{\mathcal{C}_R} \qm_k(x,E) \d x \quad.
\end{equation}
The series expansion in powers of $\hbar$ for the quantum action takes form
\begin{equation}
    \label{puffyvgn}
B(E) = S(E)
+ \dfrac{\hbar}{\iu} \sigma_1 (E)
+ \left(\dfrac{\hbar}{\iu}\right)^2 \sigma_2 (E) + \ldots \quad.
\end{equation}

Next, we substitute the expansion~\eqref{L8} into~\eqref{GBS} and get:\footnote{~As we have already mentioned, the form of the equation above implies the neglect of the tunneling effects. We will discuss this issue in Section~\ref{nonpert}.}
\begin{equation}
    B(E) = S(E) + \sum \limits_{k=1}^\infty \sigma_k(E) =
    \oint \limits_{\mathcal{C}_R} \sum \limits_{k=0}^{\infty} \left(\dfrac{\hbar}{\iu}\right)^k \qm_k(x,E) \d x
    =2\s\pi\s n\s\hbar \quad.
    \label{GBSser}
\end{equation}

The zeroth and first term in~\eqref{GBSser} give
\begin{equation}
    \oint \limits_{\mathcal{C}_R} p(x,E) \d x + \dfrac{\hbar}{\iu} \oint \limits_{\mathcal{C}_R} \qm_1(x,E) \d x =
    S(E) + \dfrac{\hbar}{\iu}\s2\s\pi\s\iu\left(-\dfrac{1}{2}\right)
    = S(E) - \pi\s\hbar
    = 2\s\pi\s n\s\hbar \quad.
    \label{derBS}
\end{equation}

The constant arising from the first term is often referred to as Maslov index and can be calculated in various ways~\cite{LL}. Importantly, it does not depend on the form of the potential well. After moving it to the RHS of~\eqref{derBS}, we arrive at the famous Bohr-Sommerfeld quantisation condition:
\begin{equation}
    S(E) = 2\s\pi\s\hbar \left(n+\dfrac{1}{2}\right)\quad.
    \label{BS}
\end{equation}

One may also proceed with calculating the higher-order terms on the LHS of~\eqref{GBSser}. This will, for example, provide a way to generate the perturbative expansion in the cases where it exists. To this end, one will have to solve~\eqref{GBSser} for the energy, inverting the series term by term.
When taking the integrals, one should take into account that all the odd terms in the expansion of the QMF starting from $k=3$ are total derivatives and therefore the corresponding integrals in~\eqref{GBSser} vanish.

\item To employ the Laurent expansion of the QMF:
\begin{equation}
    \qm(x,E) =  \sum \limits_{k=-\infty}^{\infty} a_k\s x^k\label{laurent}
\end{equation}
This approach has not been studied in depth so far. In~\cite{Nan2} and~\cite{Nan3} an approximate method of finding the energies was suggested based on this idea.
In this paper, we use the Laurent expansion to find the residue of the QMF at infinity, in order to prove the ESR property in the language of the QAF.
These results rely on those obtained in~\cite{HamJa}. We also correct some overlooks made in \textit{Ibid}.
\end{enumerate}

\section{A special property of tridiagonal matrices \label{prop}}

In this section, we prove a simple lemma interrelating the spectra of tridiagonal matrices with opposite elements on their diagonals. This lemma serves an algebraic proof of the duality of the sextic quasi-exactly solvable potentials considered in Section~\ref{algappr}.

\begin{lemma}
Let $M$ and $\widetilde{M}$ be tridiagonal matrices:
\begin{gather}
M_n = \left(
\begin{array}{cccc}
 a_1 & b_1 & 0 & 0 \\
 c_1 & a_2 & \ddots & 0\\
 0 & \ddots & \ddots & b_{n-1} \\
 0 & 0 & c_{n-1} & a_n
\end{array}\right)\quad,\quad
\widetilde{M}_n=
\left(
\begin{array}{cccc}
 -a_1 & b_1 & 0 & 0 \\
 c_1 & -a_2 & \ddots & 0\\
 0 & \ddots & \ddots & b_{n-1} \\
 0 & 0 & c_{n-1} & -a_n
\end{array}\right) \quad. \label{matr}
\end{gather}
\end{lemma}
Then their eigenvalues are equal in magnitude but opposite in sign.
\begin{proof}{}
First note that the matrices in \eqref{matr} are related through
\begin{equation}
    \widetilde{M}_n = - O_n\s M_n\s O_n \quad,
    \label{DET}
\end{equation}
where $\s O_n\s $ is an orthogonal diagonal matrix defined as
\begin{equation}
    O_{n} = \diag \{ 1,\s -1,\s 1,\s -1,\s\ldots \}\quad.
\end{equation}
Evaluation of the determinant of \eqref{DET} in the cases of even and odd $\s n\s $ gives:
\begin{equation}
\label{2}
    | \widetilde{M}_{2k-1} |= - | M_{2k-1}| \quad, \qquad
    | \widetilde{M}_{2k} |= | M_{2k} |\quad.
\end{equation}
Now let $\s P_{n}(\lambda)\s $ and $\s \widetilde{P}_{n} (\lambda)\s $ be the characteristic polynomials of the tridiagonal matrices \eqref{matr}:
\begin{gather}
    P_{n} (\lambda) = |M_{n} - \lambda \unit_n| \quad,\qquad
    \widetilde{P}_{n} (\lambda) =  |\widetilde{M}_{n} - \lambda \unit_n| \quad.
\end{gather}
If one carries out the substitution
\begin{equation}
\begin{alignedat}{6}
    &a_m &\to \phantom{~-}&a_m - \lambda \quad&\text{in } M_n\quad,\\
    -&a_m &\to ~- & a_m + \lambda \quad&\text{in } \widetilde{M}_n\quad,
\end{alignedat}
\end{equation}
then the equalities \eqref{2} turn into
\begin{gather}
    \widetilde{P}_{2\s k+1}(\lambda) = -P_{2\s k+1}(-\lambda) \quad,\qquad
    \widetilde{P}_{2\s k}(\lambda) = P_{2\s k}(-\lambda) \quad.
\end{gather}
Hence the eigenvalues of $M_n$ and $\widetilde{M}_n$ can be found from the characteristic equations
\begin{subequations}
\begin{alignat}{5}
         &M_n:& \qquad &P_{n}(\lambda) &= 0 \quad&,\qquad \\
         &\widetilde{M}_n:& \qquad &P_{n} (-\lambda) &= 0\quad&.
    \end{alignat}
\end{subequations}
Comparing the former and the latter, we observe that the eigenvalues of $\s M_n\s $ and $\s \widetilde{M}_n\s $ have opposite signs.
\end{proof}

The following detail is worth noting here.
Suppose one has derived the closed-form expressions for the roots of the $P_n(\lambda)$:
\begin{equation}\begin{aligned}
    \lambda_k = f_k (a_1, \ldots, a_n; b_1, \ldots, b_{n-1}, c_1, \ldots, c_{n-1}) \quad,\qquad k = 1\ldots n \quad.
    \label{10}
\end{aligned}\end{equation}
Generally, an individual root will not simply change its sign under the change of the signs of $a_k$:
\begin{equation}\begin{aligned}
    \widetilde{\lambda}_k =
    f_k (-a_1, \ldots, -a_n; b_1, \ldots, b_{n-1}, c_1, \ldots, c_{n-1})
    \neq - \lambda_k
    \quad,\qquad k = 1\ldots n \quad.
\end{aligned}\end{equation}
However, the roots will change the sign as a
set:
\begin{equation}
    \{ \lambda_1, \ldots, \lambda_n \} \to \{ \widetilde{\lambda_1}, \ldots, \widetilde{\lambda_n} \} =
    \{ -\lambda_1, \ldots, -\lambda_n \} \quad.
    \label{12}
\end{equation}

\section{Evaluation of residue in Ricatti equation}\label{residuesh}

In order to calculate the residue at infinity on the right-hand side of equation~\eqref{36}, we introduce the variable $y$:
\begin{subequations}
\begin{gather}
    y \equiv \dfrac{1}{x} \quad.
\end{gather}
We define the function $\tilde{\qm}\,$ as
\begin{gather}
    \widetilde{\qm}_{\{\mu\}}(y, \s -E) \equiv \qm_{\{\mu\}} (x, \s  -E) = \qm_{\{\mu\}} (1 / y, \s  -E)\quad,
\end{gather}
\end{subequations}
in whose terms the Ricatti equation~\eqref{ric2} assumes the form
\footnote{~Recall that in this section we are using the units in which $\s m=\hbar^2\s $.}
\begin{equation}
    \widetilde{\qm}^{\ 2}_{\{\mu\}}(y, \s-E) + \iu\s \hbar\s y^2 \s  \widetilde{\qm}^{\ \prime}_{\{\mu\}} (y,\s -E) = 2\s \hbar^2\s (E - V_{\{\mu\}}(1/y)) \quad.\label{ric3}
\end{equation}
The Laurent expansion for $\widetilde{\qm}_{\{\mu\}}(y,\s-E)$ becomes
\begin{equation}
    \widetilde{\qm}_{\{\mu\}}(y, \s -E) = \sum \limits_{k=-3}^{\infty} a_k\s  y^k\quad.\label{laurent0}
\end{equation}
The lowest power of $y$ is determined by the highest power of $x$ in $V(x)$\eqref{ric2}.
The insertion of the above expansion into the integral (\ref{36}) gives:
\begin{equation}
    \oint \limits_{\mathcal{C}_\infty} \qm_{\{\mu\}}(x, \s -E) \d x =
    \oint \limits_{\mathcal{C}''_0} \widetilde{\qm}_{\{\mu\}}(y, \s -E) \s \dfrac{1}{y^2}\d y = 2\s\pi\s\iu\s a_1 \quad,
\end{equation}
where the contour $\mathcal{C}''_0$ encloses the origin in the complex $y$-plane.
To find $a_1\s $, we insert the Laurent series~\eqref{laurent} in the Ricatti equation~\eqref{ric3} and match the coefficients of the five lowest powers of $y$:
\begin{equation}\begin{aligned}
k &= -6:\qquad& \dfrac{\hbar^2\s \nu^2}{4} + a_{-3}^2 &= 0 \\
k &= -5:\qquad& 2\s  a_{-3}\s a_{-2} &= 0 \\
k &= -4:\qquad& \dfrac{\hbar^2\s \nu\s \mu}{2} + a_{-2}^2 + 2\s a_{-3}\s a_{-1} &= 0 \\
k &= -3:\qquad& 2\s  (a_{-3}\s a_0 + a_{-2}\s a_{-1}) &= 0 \\
k &= -2:\qquad& \dfrac{\hbar^2\s \mu ^2}{4}-\dfrac{1}{2}\s  (8 j+3)\s \hbar^2 \s \nu +a_{-1}^2+2 a_{-2}\s  a_{0}+a_{-3} (2 a_{1}-3 \iu)&=0\quad.
\end{aligned}
\end{equation}
There are two sets of solutions:
\begin{equation}
\begin{cases}
a_{-3}= -\dfrac{\iu\s \hbar\s \nu}{2} \\
a_{-2}= 0 \\
a_{-1}= -\dfrac{\iu\s \hbar\s \mu}{2} \\
a_{0}=0\\
a_{1}= \iu \s \hbar\s (4\s  j+3)
\end{cases} \quad,\qquad
\begin{cases}
a_{-3}= \dfrac{\iu\s \hbar\s \nu}{2} \\
a_{-2}= 0 \\
a_{-1}= \dfrac{\iu\s \hbar\s \mu}{2}\\
a_{0}=0\\
a_{1}=-4 \s \iu \s \hbar\s j
\end{cases} \quad.
\end{equation}
Following~\cite{HamJa}, we employ the normalisation condition of square integrability of the wave function
\begin{equation}
\psi(x) = \exp \left[ \iu \int \limits^x \qm_{\{\mu\}}(\tilde{x}, \s -E) \d{} \tilde{x} \right]\quad.
\end{equation}
For large $x$ (and small $y$) we get:
\begin{equation}
\psi(x) \approx \exp \left[ \iu\; \dfrac{a_{-3}\s x^4}{4}\s  \right] \quad,
\end{equation}
and we have to choose the second set of solutions with $a_{-3} = \iu\s \hbar\s \nu/2$ and $a_1 = - 4 \s \iu\s \hbar\s j$. Thus we arrive at
\begin{equation}
    2\s \pi\s \iu\s \Res\{\qm_{\{\mu\}} (x, \s -E), \infty\}
    = 2\s \pi\s \iu\s a_1 = 8 \s  \pi \s j\s \hbar \quad.
\end{equation}

\section{Possible shapes of sextic QES potentials \label{shapes}}


We are interested in the following special cases of potential~\eqref{Vcl}:
\begin{enumerate}
\newcounter{potcases}
\item In the case of
\begin{equation}
\addtocounter{equation}{1}
    b < 0\quad,
    \addtocounter{potcases}{1}
    \tag{\theequation\alph{potcases}}
    \label{case1}
\end{equation}
the potential has a double-well shape, with two absolute minima and one local maximum.
\item In the case of
\begin{equation}
    0 < b < \dfrac{c^2}{4\s d}\quad,
    \addtocounter{potcases}{1}
    \tag{\theequation\alph{potcases}}
    \label{case2}
\end{equation}
the following two subcases will be addressed:
\begin{itemize}
\item If
\begin{equation}
    c > 0 \quad,
    \addtocounter{potcases}{1}
    \tag{\theequation\alph{potcases}}
    \label{case21}
\end{equation}
the potential is of a single-well type, i.e., has a single absolute minimum and no maxima;
\item If
\begin{equation}
    c<0\quad,
    \addtocounter{potcases}{1}
    \tag{\theequation\alph{potcases}}
    \label{case22}
\end{equation}
the potential has three wells, and has one of the local minima located at $\s x=0\s $.
\end{itemize}
\end{enumerate}
The case of $\s b>\dfrac{ c^2}{ 4\s d}\;$ is of no interest to us and will not be considered here.

For different values of the parameters $\mu$, $\nu$ and $j$, the QES sextic potentials~\eqref{V} fall either into the category~\eqref{case1} or into~\eqref{case2}. For both categories, the corresponding classical systems possess the duality property. Below we shall limit our discussion to the even potential.

Dual to one another are two potentials with the opposite values of $\mu$. We intend to show that there always exists a range of energies, for which at least one of the two potentials has four solutions, while the other one has two.
\begin{enumerate}
\item In the case of
\begin{equation}
    \dfrac{\mu^2}{8} < 2\s \nu\s \left(j+\dfrac{3}{8}\right)\quad,
    \label{qcond1}
\end{equation}
both potentials, $V_{\{\mu\}}(x)$ and $V_{\{-\mu\}}(x)$ are of a double-well shape, see Figure~\ref{F1}, (a) and (b). The duality takes place in the following range of energies:
\begin{equation}
\begin{gathered}
\text{Two wells} \hspace{45mm} \text{Single well} \hspace{45mm} \\
\begin{alignedat}{14}
    &V_{\{\mu\},\s\text{min}} &<& \ E < V_{\{\mu\}}(0) \qquad \qquad &V_{\{\mu\}}(0) &<& \ E  &<& -&V_{\{-\mu\},\s\text{min}}
    &\quad\quad\qquad &\text{(for $V_{\{\mu\}}$)} \quad&,\\
    &V_{\{-\mu\},\s\text{min}} &<& \ E < V_{\{-\mu\}}(0) \qquad \qquad &V_{\{-\mu\}}(0) &<& \ E  &<& -&V_{\{\mu\},\s\text{min}}
    &\quad\quad\qquad &\text{(for $V_{\{-\mu\}}$)} \quad&.
\end{alignedat}
\end{gathered}
\label{zho1}
\end{equation}
\item In the case of
\begin{equation}
    \dfrac{\mu^2}{8} > 2\s \nu\s \left(j+\dfrac{3}{8}\right)\quad,
    \label{qcond2}
\end{equation}
one of the potentials has a single minimum while the other one has three of them, see Figure~\ref{F1}, (c) and (d). For definiteness, assume $\mu < 0$. Then the energy ranges wherein the duality occurs look as follows:
\begin{equation}
\begin{gathered}
\text{Two wells} \hspace{45mm} \text{Single well} \hspace{45mm} \\
\begin{alignedat}{14}
    &V_{\{\mu\},\s\text{min}} &<& \ E < V_{\{\mu\}}(0) \qquad \qquad
    & & & & & &
    &\quad\quad\qquad &\text{(for $V_{\{\mu\}}$)} \quad&,\\
    & & &
    \qquad \qquad &V_{\{-\mu\}}(0) &<& \ E  &<& -&V_{\{\mu\},\s\text{min}}
    &\quad\quad\qquad &\text{(for $V_{\{-\mu\}}$)} \quad&,
\end{alignedat}
\end{gathered}
\label{zho2}
\end{equation}
\end{enumerate}
In both cases, $V_{\{\mu\},\s\text{min}}$ and $V_{\{-\mu\},\s\text{min}}$ denote the values of the corresponding potentials at the minima, and we have the following equality:
\begin{equation}
    V_{\{\mu\}}(0) = - V_{\{-\mu\}}(0) \quad.
\end{equation}
From the formul\ae~\eqref{zho1} and~\eqref{zho2} it ensues that, without loss of generality, we can assume that the equation
\begin{equation}
     V_{\{\mu\}}(x) = -E
     \label{Veps}
\end{equation}
has four solutions. Consequently, the equation
\begin{equation}
     V_{\{-\mu\}}(x) = E
\end{equation}
will then have two real solutions.

For the odd potential, one has to replace $3/8$ with $5/8$ in both~\eqref{qcond1} and~\eqref{qcond2}.

\section{Action in potentials with multiple minima \label{quaction}}

Equation~\eqref{solving} reveals a peculiarity inherent in all potentials with degenerate minima.
For such potentials, the set of integer numbers on the RHS of~\eqref{GBS}, for which the GBS has solutions, typically has gaps in it. In the case of the double-well potential the set is
\begin{equation}
    n \in \{0,\s 1,\s 2,\s\ldots,\s m^*-1,\s m^*,\s k^*,\s k^*+1,\s k^*+2,\s\ldots\} \quad.
\end{equation}
Here $m^*$ denotes the largest value of $n$ for which~\eqref{GBS} has solutions (or a solution) below its maximum, while $k^*$ is the smallest one for which the solution above the maximum exists.
This, however, should not surprise us.
Indeed, since the LHS of~\eqref{GBS} has the meaning of action, it should experience a jump, as the energy crosses the level of the maximum of the potential. For symmetric double-well potentials, we can expect
\begin{equation}
    2\s m^* \approx k^* \quad.
    \label{app}
\end{equation}
In general, the above equality is not exact~---~though it becomes asymptotically exact for highly excited states (i.e., for $\s m^* \to \infty$).
In Appendix~\ref{exampleee}, this will be illustrated with examples.

\section{Numbering of energy levels. Examples \label{exampleee}}

Here we study in detail the numbering of levels of the even potential~\eqref{even} with $j=1$ (i.e., $N = 4$) and $\nu = 1$, and with different values of $\mu$. Let $m_{\{\mu\},\s n}$ be such a value of the integer $\s m\s $ in
\begin{equation}
    B_{\{\mu\}}(E) = \oint \limits_{\mathcal{C}_R} \qm_{\{\mu\}}(x, E) \d x = 2 \s  \pi \s m\s \hbar\s \quad,\label{GBSm}
\end{equation}
that the solution of this equation gives the energy level $E_{\{\mu\},\s n}$.
As we discussed above, in the case of a single well it is but the number of the energy level, $m_{\{\mu\},\s n} = n$. In the case of two wells, one either has two energy levels with numbers $n=2\s m_{\{\mu\},\s n}$ and $n=2\s m_{\{\mu\},\s n}+1$ or, possibly, solely the level $n=2\s m_{\{\mu\},\s n}$. Due to the ESR property, these numbers for the dual potentials are linked via
\begin{equation}
    \left[
    \begin{alignedat}{4}
    &m_{\{\mu\},\s n} + 2\s m_{\{-\mu\},\s N-n} & \ = \ & N\\
    &2 \s m_{\{\mu\},\s n} + m_{\{-\mu\},\s N-n} & \ = \ & N
    \end{alignedat}
    \right.
    \quad,\qquad
    n=0,2,4\ldots,N
    \quad.
\end{equation}
In other words, if one of two symmetric levels lies below the maximum of the potential, another one has to reside above the maximum.

\begin{enumerate}
\item $\mu = 5$

In this case, the potential $V_{\{\mu\}}(x)$ has single-well shape, while $V_{\{-\mu\}}(x)$ is of a triple-well shape. The numbering of the levels is
\begin{equation}
\begin{tabular}{|c|c|c|}
\hline
\multicolumn{3}{|c|}{$\mu = 5$} \\
\hline
    $n$ & $m_{\{\mu\},\s n}$ & $m_{\{-\mu\},\s n}$ \\
\hline
    5 & 5 & 2 \\
    4 & 4 & 2 \\
    3 & 3 & 1 \\
    2 & 2 & 1 \\
    1 & 1 & 0 \\
    0 & 0 & 0 \\
\hline
\end{tabular}
\label{tabu5}
\end{equation}
In the right column, one can clearly observe the pattern usual for  energy splitting.
\begin{subequations}
\begin{alignat}{6}
m_{\{\mu\},\s 0} + 2\s m_{\{-\mu\},\s 4} = 0 + 2 \cdot 2 = 4 \quad,\label{sums51}\\
m_{\{\mu\},\s 2} + 2\s m_{\{-\mu\},\s 2} = 2 + 2 \cdot 1 = 4 \quad,\\
m_{\{\mu\},\s 4} + 2\s m_{\{-\mu\},\s 0} = 4 + 2 \cdot 0 = 4 \quad.
\end{alignat}
\label{sums5}
\end{subequations}

\item $\mu = 4$

In this case, both potentials are of a double-well shape. However, the ground state energy in $V_{\{\mu\}}(x)$ is above the maximum of the potential:
\begin{equation}
    E_{\{\mu\},\s 0} > V_{\{\mu\}}(0) \quad,
\end{equation}
and the numbering of the levels is the same as in~\eqref{tabu5}.

\item $\mu = 3$

In this case, the ground state energy in $V_{\{\mu\}}(x)$ is still above the maximum of the potential. However, the numbering of the levels changes slightly (though this change does not touch the levels possessing the ESR property).
\begin{equation}
\begin{tabular}{|c|c|c|}
\hline
\multicolumn{3}{|c|}{$\mu = 3$} \\
\hline
    $n$ & $m_{\{\mu\},\s n}$ & $m_{\{-\mu\},\s n}$ \\
\hline
    5 & 5 & 5 \\
    4 & 4 & 2 \\
    3 & 3 & 1 \\
    2 & 2 & 1 \\
    1 & 1 & 0 \\
    0 & 0 & 0 \\
\hline
\end{tabular}
\label{tabu3}
\end{equation}
The fifth energy level in the potential $V_{\{-\mu\}}(x)$ is no longer below the maximum of the potential. We can summarise the difference as follows:
\begin{equation}
\begin{alignedat}{6}
    &\mu = 4\text{:} \qquad m_{\{-\mu\},\s 5} &= 2 \quad,\\
    &\mu = 3\text{:} \qquad m_{\{-\mu\},\s 5} &= 5 \quad.
\end{alignedat}
\end{equation}
In other words, for the potential $V_{\{-4\}}(x)$, the fifth energy level can be found from the equation
\begin{equation}
    \mu = 4\text{:} \qquad
    \oint \limits_{\mathcal{C}_R} \qm_{\{-\mu\}}(x, E) \d x = 2\s \pi\s (2)\s \hbar \qquad \to \qquad E_{\{-\mu\},\s 5} \quad,
\end{equation}
while for the potential $V_{\{-3\}}(x)$, the fifth energy level is obtained from the equation
\begin{equation}
    \mu = 3\text{:} \qquad
    \oint \limits_{\mathcal{C}_R} \qm_{\{-\mu\}}(x, E) \d x = 2\s \pi\s (5)\s \hbar \qquad \to \qquad E_{\{-\mu\},\s 5} \quad.
\end{equation}

\item $\mu = 2.5$

In this case, the ground state energy in the potential $V_{\{\mu\}}(x)$ is below its maximum, while the first excited state is above the maximum.
This is sufficient for us to state that the fourth state of the potential $V_{\{-\mu\}}(x)$ is above its maximum.
\begin{equation}
\begin{tabular}{|c|c|c|}
\hline
\multicolumn{3}{|c|}{$\mu = 2.5$} \\
\hline
    $n$ & $m_{\{\mu\},\s n}$ & $m_{\{-\mu\},\s n}$ \\
\hline
    5 & 5 & 5 \\
    4 & 4 & 4 \\
    3 & 3 & 1 \\
    2 & 2 & 1 \\
    1 & 1 & 0 \\
    0 & 0 & 0 \\
\hline
\end{tabular}
\label{tabu25}
\end{equation}
\begin{subequations}
\begin{alignat}{6}
&2\s m_{\{\mu\},\s 0} + m_{\{-\mu\},\s 4} & \ = \ & 2 \cdot 0 + 4 & \ = \ & 4 \quad,\label{sums251}\\
&m_{\{\mu\},\s 2} + 2\s m_{\{-\mu\},\s 2} & \ = \ & 2 + 2 \cdot 1 & \ = \ & 4 \quad,\\
&m_{\{\mu\},\s 4} + 2\s m_{\{-\mu\},\s 0} & \ = \ & 4 + 2 \cdot 0 & \ = \ & 4 \quad.
\end{alignat}
\label{sums25}
\end{subequations}

Mind the difference between~\eqref{sums51} and~\eqref{sums251}. We had to add the factor of two to $m_{\{\mu\},\s 0}$ due to the fact that the ground state is below $V_{\{\mu\}}(0)$ for $\mu = 2.5$.

One may have noticed from~\eqref{tabu25} that, as we go from the third to the fourth energy level in $V_{\{-\mu\}}(x)$, the quantum action becomes four times larger. This is, of course, a purely quantum effect. The general rule for our potential is:
\begin{equation}
    B(E_{\{\mu,\s k^*\}}) - 2\s  B(E_{\{\mu,\s m^*\}}) \leq 4 \s  \pi \s  \hbar \quad,
\end{equation}
with $k^*$ and $m^*$ introduced in Appendix~\ref{quaction}.

\item $\mu = 2$

In this case, for $V_{\{\mu\}}(x)$ one has two levels below its maximum, and has four levels below the maximum for $V_{\{-\mu\}}(x)$.
\begin{equation}
\begin{tabular}{|c|c|c|}
\hline
\multicolumn{3}{|c|}{$\mu = 2$} \\
\hline
    $n$ & $m_{\{\mu\},\s n}$ & $m_{\{-\mu\},\s n}$ \\
\hline
    5 & 5 & 5 \\
    4 & 4 & 4 \\
    3 & 3 & 1 \\
    2 & 2 & 1 \\
    1 & 0 & 0 \\
    0 & 0 & 0 \\
\hline
\end{tabular}
\label{tabu2}
\end{equation}
The enumeration of the energy levels possessing the ESR property has not changed, and stays same as in~\eqref{sums25}.

\item $\mu = 0$

In this case, the self-duality takes place, see Eq.~\eqref{selfd2}.
\begin{equation}
\begin{tabular}{|c|c|c|}
\hline
\multicolumn{3}{|c|}{$\mu = 0$} \\
\hline
    $n$ & $m_{\{\mu\},\s n}$ & $m_{\{-\mu\},\s n}$ \\
\hline
    5 & 5 & 5 \\
    4 & 4 & 4 \\
    3 & 3 & 3 \\
    2 & 1 & 1 \\
    1 & 0 & 0 \\
    0 & 0 & 0 \\
\hline
\end{tabular}
\label{tabu0}
\end{equation}
The energy of the second excited state is equal precisely to the value of the potential at the maximum, which is a general rule for integer values of $j$.
\begin{equation}
    \begin{cases}
    \mu = 0 \\
    j \in \mathbb{N}
    \end{cases}
    \qquad \to \qquad
    E_{\{\mu\},\s N/2} = V_{\{\mu\}}(0) = 0 \quad.
\end{equation}
This energy level can be treated both as (a)~a level below the maximum, with zero non-perturbative splitting; or (b)~a level above the maximum, with a zero non-perturbative shift. The summation reads as:
\begin{subequations}
\begin{align}
    & \hspace{3.5mm}         2\s  m_{\{\mu\},\s 2} + 2\s m_{\{-\mu\},\s 2} =  2 \cdot 1           +  2 \cdot 1 = 4 \hspace{1.315em},\\
    &\left[
    \begin{aligned}
    & 2\s m_{\{\mu\},\s 2} + 2\s m_{\{-\mu\},\s 2} = 2 \cdot 1  +  2 \cdot 1 =  4 \\
    & \phantom{2\s }m_{\{\mu\},\s 2}  +  2\s m_{\{-\mu\},\s 2} = \hspace{5.2mm} 2  +  2 \cdot 1  =  4  \\
    & 2\s m_{\{\mu\},\s 2}  + \phantom{2\s }m_{\{-\mu\},\s 2} =  2 \cdot 1  +  \hspace{5.2mm} 2 =  4 \\
    & \phantom{2\s }m_{\{\mu\},\s 2}  +  \phantom{2\s }m_{\{-\mu\},\s 2} = \hspace{5.2mm} 2  +  \hspace{5.2mm} 2  = 4
    \end{aligned}
    \right.\quad,\\
    &\hspace{3.5mm} \phantom{2\s }m_{\{\mu\},\s 4} + 2\s m_{\{-\mu\},\s 0} =  \hspace{5.2mm} 4 + 2  \cdot 0 = 4 \hspace{1.315em}.
\end{align}
\end{subequations}
\end{enumerate}

\section{Discussion of results by Geojo, Ranjani and Kapoor}

As we have already mentioned, our studies of the QES systems in the language of the QAF formalism  were largely inspired by the results obtained in~\cite{HamJa} by Geojo, Ranjani and Kapoor. We agree with {\it{Ibid.}} on the following items:
\begin{itemize}
    \item For the sextic potential, the assumption of the QMF to have an isolated pole at infinity leads to the condition of quasi-exact solvability for this potential.
    \item The residue of the QMF can be calculated explicitly by substituting the Laurent expansion of the QMF into the Ricatti equation.
\end{itemize}
However, in~\cite{HamJa} an oversight was made when the authors implicitly assumed the contour enclosing the real poles of the QMF to enclose \textit{all} the poles of the QMF. In our notation (see Figure~\ref{roots6}), this oversight reads as:
\begin{equation}
    \xcancel{ \oint \limits_{\mathcal{C}_3} \qm_{\{\mu\}}(x, E) \d x = \oint \limits_{\mathcal{C}_\infty} \qm_{\{\mu\}}(x, E) \d x} \quad.
\end{equation}
In reality, however, the singularities which are off the real axis should also be taken into account, in order to obtain the correct expression~\eqref{36}.

Due to the said omission, the authors got a wrong result for the residue of the QMF at infinity, which in their case turned out to be dependent on the number of the energy level.$\s $\footnote{~Specifically, the meaning of $n$ from equation (6) in~\cite{HamJa} had to be different from the meaning of $n$ from equation (20) in {\it{Ibid}}. In the former equation, $\s n\s $ is related to the number of the energy level, while in the latter it is the number of the highest energy level from the algebraic sector ($N$ in our notation).}

\end{appendices}

 \end{document}